\documentclass[preprint2,twocolappendix]{aastex6}

\usepackage{color}
\usepackage[normalem]{ulem} 
\usepackage{soul} 

\slugcomment{To appear in Astrophysical J., 833:45.}

\shorttitle{Effect of coronal pressure waves}
\shortauthors{Rouillard et al.}

\begin{document}


\title{Deriving the properties of coronal pressure fronts in 3-D: application to the 17 May 2012 ground level enhancement}


\author{A. P. Rouillard\altaffilmark{1,2}, I. Plotnikov\altaffilmark{1,2}, R.F. Pinto\altaffilmark{1,2},  M. Tirole\altaffilmark{1,2}, M. Lavarra\altaffilmark{1,2}} 
\author{P. Zucca\altaffilmark{3}} 
\author{R. Vainio\altaffilmark{4}} 
\author{A.J. Tylka\altaffilmark{5}}
\author{A. Vourlidas\altaffilmark{6}}
\author{M.L. De Rosa\altaffilmark{7}} 
\author{J. Linker\altaffilmark{8}} 
\author{A. Warmuth\altaffilmark{9}, G. Mann\altaffilmark{9}} 
\and \author{C.M.S. Cohen\altaffilmark{10}, R.A. Mewaldt\altaffilmark{10}} 
\email{arouillard@irap.omp.eu}


\altaffiltext{1}{Institut de Recherche en Astrophysique et Plan\'{e}tologie, Universit\'{e} de Toulouse III (UPS), France}
\altaffiltext{2}{Centre National de la Recherche Scientifique, UMR 5277, Toulouse, France}
\altaffiltext{3}{LESIA-UMR 8109 - Observatoire de Paris, CNRS, Univ. Paris 6 $\&$ 7, 92190, Meudon, France}
\altaffiltext{4}{University of Turku, Turku, Finland}
\altaffiltext{5}{Emeritus, NASA Goddard Space Flight Center, Greenbelt, Maryland, SUA}
\altaffiltext{6}{Johns Hopkins Applied Physics Laboratory, Laurel, Maryland, USA}
\altaffiltext{7}{Lockheed Martin Solar and Astrophysics Laboratory, Palo Alto, California, USA}
\altaffiltext{8}{Predictive Sciences Inc., San Diego, California, USA}
\altaffiltext{9}{Leibniz-Institut f\"{u}r Astrophysik Potsdam (AIP), Potsdam, Germany}
\altaffiltext{10}{California Institute of Technology, Pasadena, California, USA}


\begin{abstract}

We study the link between an expanding coronal shock and the energetic particles measured near Earth during the Ground Level Enhancement (GLE) of 17 May 2012. We developed a new technique based on multipoint imaging to triangulate the 3-D expansion of the shock forming in the corona. It uses images from three vantage points by mapping the outermost extent of the coronal region perturbed by the pressure front. We derive for the first time the 3-D velocity vector and the distribution of Mach numbers, $M_{FM}$, of the entire front as a function of time. Our approach uses magnetic field reconstructions of the coronal field, full magneto-hydrodynamic simulations and imaging inversion techniques. We find that the highest $M_{FM}$ values appear near the coronal neutral line within a few minutes of the Coronal Mass Ejection (CME) onset; this neutral line is usually associated with the source of the heliospheric current and plasma sheet. We illustrate the variability of the shock speed, shock geometry and Mach number along different modeled magnetic field lines. Despite the level of uncertainty in deriving the shock Mach numbers, all employed reconstruction techniques show that the release time of GeV particles occurs when the coronal shock becomes super-critical ($M_{FM}>3$). Combining in-situ measurements with heliospheric imagery, we also demonstrate that magnetic connectivity between the accelerator (the coronal shock of 17 May 2012) and the near-Earth environment is established via a magnetic cloud that erupted from the same active region roughly five days earlier.   

\end{abstract}



\section{Introduction}

\indent 
The link between Coronal Mass Ejection, the perturbations of the corona they induce and the production of Solar Energetic Particles (SEPs) is a topic of active research. During the launch of an energetic CME, moving fronts, or waves, are frequently observed in Extreme UltraViolet images (EUV) propagating away from the flaring source region (Thompson et al. 1999). There is a great event to event variability in the morphology and kinematic properties of these EUV fronts making their physical interpretation challenging. For a comprehensive discussion of all proposed theories concerning their origins, we here refer the reader to the extensive review by Warmuth (2015) on this topic. In addition, the formation of the White-Light (WL) signatures of CME-driven shocks was investigated observationally by Ontiveros and Vourlidas (2009) (see also review by Vourlidas and Ontiveros 2010) and numerically by Manchester al. (2008). The \emph{Sun-Earth Connection Coronal and Heliospheric Investigation} (\emph{SECCHI}; Howard et al. 2008) onboard the Solar-Terrestrial Relation Observatory (STEREO) mission (Kaiser et al. 2008) has provided since 2007, unprecedented imaging of solar storms from vantage points situated outside the Sun-Earth line. This capability combined with the images taken by the \emph{Atmospheric Imaging Assembly} (AIA) onboard the Solar Dynamics Observatory (SDO) (Lemen et al. 2012) has spurred a flurry of studies on Magnetic Flux Ropes (MFRs) and coronal pressure waves that form during CME events.\\

\indent   The so-called 3-part structure of a CME often observed in WL images includes a filament, a dark core and a pile-up. The pile-up marks initially the outer contour of a CME (e.g. Hundhausen et al. 1972, see review by Thernisien et al. 2011); it corresponds to plasma lifted from the low corona and/or pushed aside by the dark core where the MFR acts as an expanding piston (Vourlidas et al. 2013). Remote-sensing observations combined with numerical simulations show that the subset of EUV fronts that form at the coronal base during CME onset is initially co-located with the 'pile-up' and corresponds to material compressed at low coronal heights by the lateral expansion of the flux rope  (e.g. Patsourakos and Vourlidas 2009; Rouillard et al. 2012).  When the lateral expansion ceases because the core has reached some pressure equilibrium with the surrounding coronal medium, it can no longer push material in the low corona along the surface and the EUV wave gradually becomes more freely propagating (Patsourakos and Vourlidas 2012; Warmuth, 2015). Its speed and direction are no longer dictated by the expanding core but gradually becomes altered by the local variations in the characteristic speed of the medium. This propagation phase was studied in a number of papers that, not only tracked the EUV signatures, but also the induced deflection of coronal material higher up in the corona (Rouillard et al. 2012; Kwoon et al. 2015).  \\

\indent Fast CMEs form near active regions that are typically situated below helmet streamers where strong magnetic fields can prevail. In the direct vicinity of active regions, the characteristic speed of plasma can reach values greater than 1000 km/s but EUV front speeds are typically less than 1000 km/s (Nitta et al. 2013), hence many EUV fronts may not have enough time to steepen into shocks near active region. The EUV front could be initially a layer of compressed material separating the MFR with the ambient corona plasma. It is only when the ambient characteristic speed has sufficiently decreased away from the source region, that a fast pressure front driven by the expansion of the MFR may eventually steepen into a shock. It is impossible to tell from EUV or WL images alone if a shock has really formed at a particular height and a technique must be developed to infer where the propagating front moves faster that the local fast-mode speed . This is one of the challenging tasks undertaken in this paper.\\ 

\indent We also investigate here the relation between the evolving CME and the release of high-energy particles near the Sun on 17 May 2012. The physical mechanisms that produce solar particles with energies greater than several 100 MeV within a few minutes of the flare and/or CME occurrence are still highly debated. Different origins have been proposed, including magnetic reconnection in solar flares (e.g. Cane et al. 2003), betatron acceleration in the interaction region generated by the expanding CME (e.g. Kozarev et al. 2013), diffusive shock acceleration in the  shock located around the rapidly expanding CME (e.g. Sandroos and Vainio 2009). Recent studies have exploited the unprecedented imaging capability offered by STEREO and SDO to track and compare the 3-D evolution of propagating fronts with the properties of SEPs near 1AU (Rouillard et al. 2012, Lario et al. 2014, Kozarev et al. 2015). To do that, the propagation time required for particles to reach the spacecraft making in-situ measurements must be accounted for by considering both their transit speed and the distance travelled. The latter is regulated by the length and variability of the interplanetary magnetic field. These very few studies show that the timing of SEP onsets can be understood in terms of the time taken by the fast coronal shock to reach the different magnetic field lines connected with particle detectors. The questions that remain unanswered are: (1) where along those fronts does the shock form and, (2) is the shock sufficiently strong in the corona to energise particles?

\indent The analysis of the 21 March 2011 event by Rouillard et al. (2012) showed that the 30 minute delay of the two onsets of SEP events measured at L1 relative to STEREO-A (STA) was the time for the propagating front to transit from the footpoint location of the magnetic field lines connected with STA to those connected with the L1 spacecraft. For the reasons discussed in the previous sections, testing the hypothesis that particles are accelerated at the CME shock cannot be limited to simply tracking propagating fronts in EUV images. Hence Rouillard et al. (2012), presented a combined analysis of the EUV and WL corona to derive an estimate of the 3-D speed of the pressure wave by tracking both the density variations ahead of the CME and deflected streamers higher up in the corona. No derivation was proposed in that study of the fast magnetosonic Mach number that would confirm the existence of a shock at any particular height. However as we shall see in this paper, the coronal heights considered in Rouillard et al. (2012) were likely high enough for the ambient fast mode speed to have dropped sufficiently for a shock to form. Since this study, we have developed a number of observationally-based techniques to derive quantitatively the 3-D properties of propagating fronts (including the $M_{FM}$) in order to test the hypothesis that high-energy SEP are produced at coronal shocks. \\

\indent 
After presenting the properties of the 17 May 2012 event (sections 2, 3 and 4), we present a new method to extract shock wave parameters in 3-D (sections \ref{sec:PFSS} and \ref{sec:MHD} ) using a number of different techniques. We then compare those derived shock parameters with simultaneous radio measurements (section 7) and the properties of the SEP measured near Earth (section 8).

\section{The 17 May SEP event:}

\indent At  01:25UT on 17 May 2012, the Geostationary Operational Environmental Satellites (GOES) spacecraft detected a M5.1 X-ray flare after several days of relatively quiet solar activity marked by occasional C-class flares, weak CME events ($<$600 km/s) and relatively weak energetic particle fluxes measured in the inner heliosphere. This M-class flare was associated with the eruption of a fast ($>$1600 km/s) and impulsive CME and the detection of very energetic particles (GeV) near Earth. A previous study reported that this solar event was associated with the detection of a Ground Level Enhancement (GLE) by ground-based neutron monitors (Gopalswamy et al., 2013); evidence that proton exceeding several hundreds of MeV energies were released from the Sun. This is directly supported by space measurements of protons exceeding GeV energies (Adriani et al. 2015) by the The Payload for Antimatter Matter Exploration and Light-nuclei Astrophysics (PAMELA) instrument (Picozza et al. 2007). This event occured in isolation provides an excellent opportunity to study the link between a CME and the production of high-energy particles without the contamination from other events.\\ 

\indent There are several puzzling aspects of this event that were highlighted in previous articles. In particular the flare intensity (M5.1) was lower than flare intensities measured in previous GLE events. As noted by Gopalswamy et al. (2013) and discussed in detail later in this paper; despite the rather weak flare, the associated CME had a fast speed more typical of X-class flares. Based on the arrival time of the GeV particles detected in the GLE and an interplanetary magnetic field line of length 1.2AU, Gopalswamy et al. (2013) put the Solar Particle Release (SPR) time of particles near the Sun at 01:41UT or about 15 minutes after flare onset when the CME had already reached a height of 2.3 R$_\odot$ and roughly ten minutes after the onset time of the type II burst (01:30UT). Using simple geometric arguments, they put the height of a first shock formation roughly at 1.38 R$_\odot$ well below the height reached by the leading edge of the CME at their SPR time. In Appendix A, we use a velocity dispersion analysis to show that the SPR time derived by Gopalswamy is likely too late by some 4 minutes (01:37:20$\pm$00:00:02UT) because the pathlength followed by these particles is more likely to be about 1.89$\pm0.02$ AU. Provided that magnetic connectivity between the shock and the point of in-situ measurements is maintained from the time of shock formation onwards the shock would have about five minutes to accelerate particles to GeV energies. The hypothesis that diffusive-shock acceleration is the energisation mechanism of these particles assumes that magnetic connectivity is established between the Earth and a coronal shock. This paper presents a thorough analysis of the evolution of the shock and employs a new combination of observationally-based and numerical techniques to derive, not only the magnetic connectivity of the near-Earth environment with the shock, but also some of the shock properties before and during the GLE event.\\

\section{Observations:}

\begin{figure}
\epsscale{1.1}
\plotone{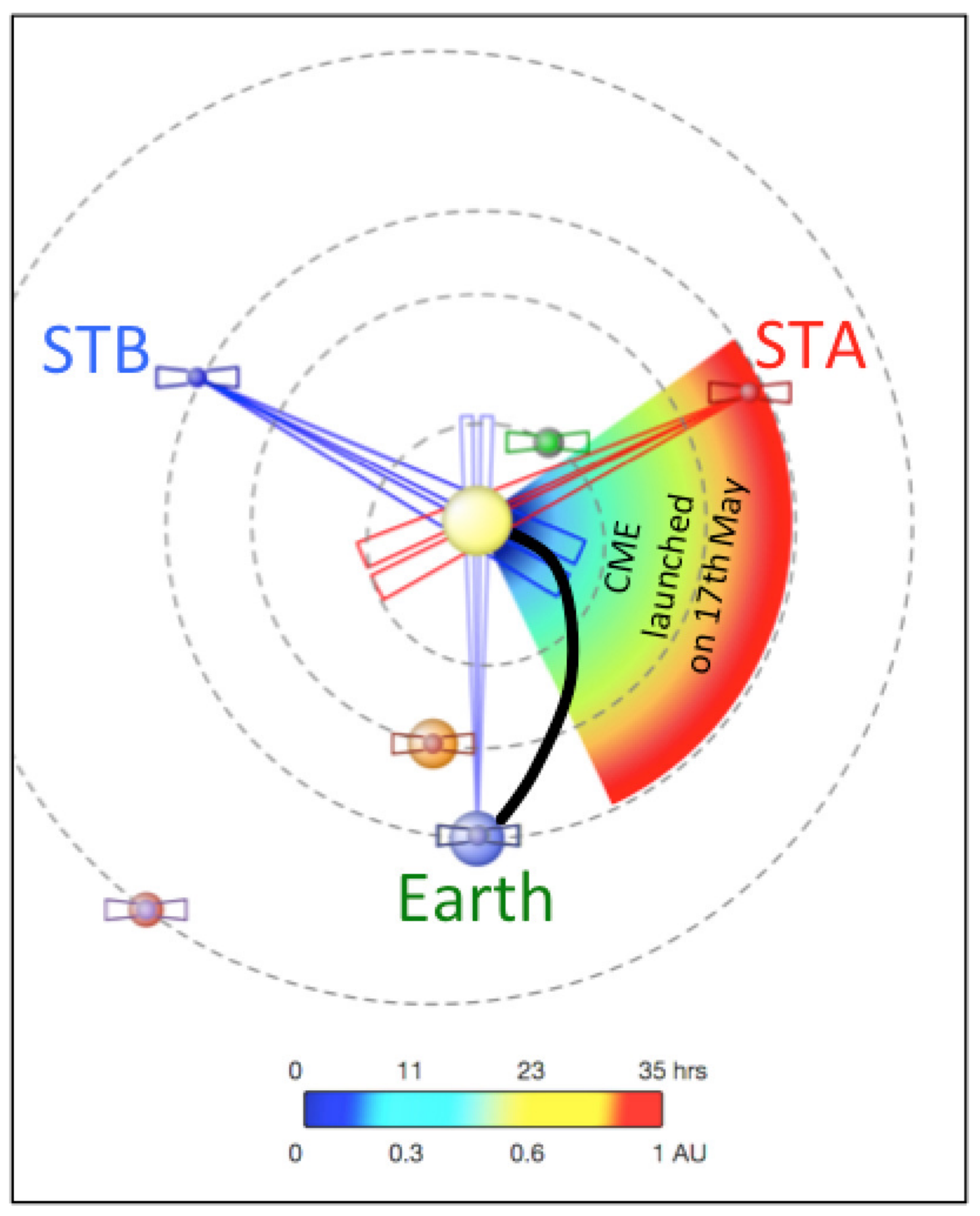}
\caption{ A view of the ecliptic plane from solar north showing the positions of the Earth, STA and STB. The nominal Parker spiral connecting magnetically the Earth to the low corona is shown in black. The intersection of the COR2A (red), COR2-B (dark blue), SOHO C2 (light blue) fields of views with the ecliptic plane are shown as pairs of elongated triangles. The trajectory of the CME launched on the 17th of May results from the analyses of heliospheric imagery as given in Appendix A). The longitudinal extent of the CME (piston+shock) was chosen to fit with the observation of the shock by STA (as measured in situ: see Appendix A) and is here exactly 100 degrees. This figure and the analysis of the trajectory of the CME was made using the IRAP propagation tool and J-maps produced by the HELCATS project (see acknowledgements for details).  {\it The Astrophysical Journal}}
\label{ORBITSEP}
\end{figure}

Figure~\ref{ORBITSEP}  presents the positions of the STEREO spacecraft and the Earth on 17 May 2012, these three vantage points provided 360$^\circ$ views of the Sun. The longitudinal separation of STA and STB with respect to Earth were 114$^\circ$ and 117$^\circ$, respectively. The expansion of the CME could be tracked simultaneously from widely separated spacecraft allowing the 3-D volume of the expanding high-pressure fronts to be derived by using the comprehensive suite of optical instruments on STEREO, SDO and the \emph{Solar and Heliospheric Observatory} (\emph{SOHO}). The SECCHI package onboard \emph{STEREO} (Howard et al. 2008) consists of an Extreme Ultraviolet Imager (EUVI), two coronagraphs (COR-1 and COR-2), and the Heliospheric Imager (HI).  At the time of the event studied here, the magnetic connectivity of the STEREO and the near-Earth orbiting spacecraft, also provides a circumsolar measurement of particles potentially released from widely separated source regions.      \\

\begin{figure*}
\begin{center}
\includegraphics[angle=0,scale=.35]{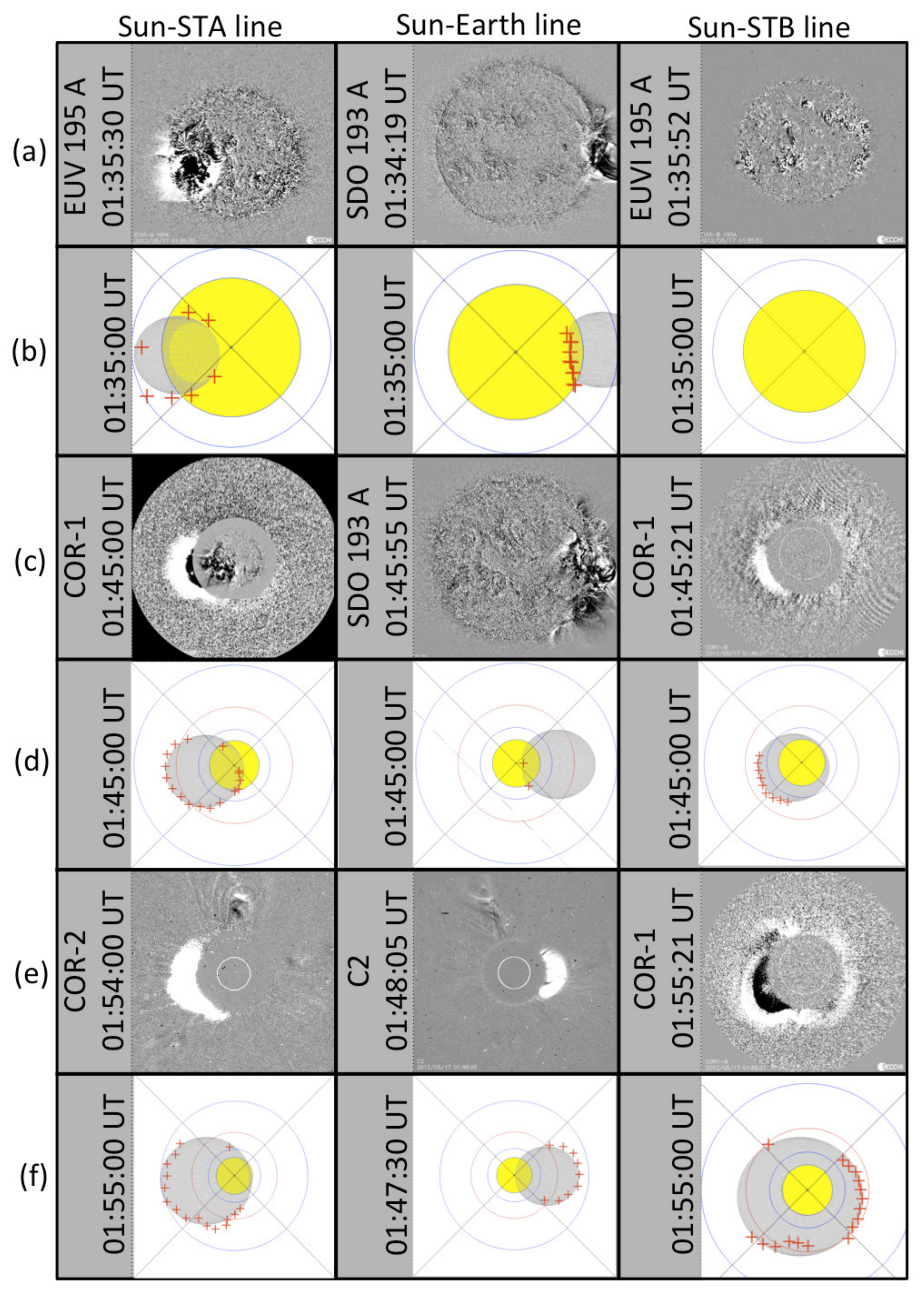}
\caption{This figure compares running-difference images (rows a,c,e) of the CME observed by STA (left hand-column and STB (right-hand column) with the results of applying the fitting technique (rows b,d,f) developed here. The images are all from the EUVI instruments except the left-hand image shown in row f obtained by COR1-A. Red crosses are superposed on the fitted ellipsoids, they show the contour of the propagating front observed in the running difference images and are used to constrain the extent and location of the ellipsoid at each time.}
\label{TRIANGWL}
\end{center}
\end{figure*}

\indent Rows (a), (c) and (e) of Figure~\ref{TRIANGWL} present images covering the first 20 minutes of the CME eruption as viewed along the Sun-STA, Sun-Earth and Sun-STB lines. With the exception of the image obtained by the \emph{Large Angle and Spectrometric Coronagraph Experiment} (\emph{LASCO}; Brueckner et al. 1995) C2 shown in the last row, the sequence of images shown in rows (a), (c) and (e) are at the closest times to 01:35, 01:45 and 01:55UT, respectively. These images show that the coronal region perturbed by the expanding CME increases with time in both EUV and in the WL images. \\

\indent The surface of the propagating front generated around the expanding CME is initially fairly regular and we found that an ellipsoid fits the outermost extent of this perturbed region very well. We manually extracted the location of the outermost extent of the CME off limb and on disk at all available times. These points are plotted as red crosses in the images given in rows (b), (d) and (f) and are used to outline the contour of the ellipsoids viewed from the three vantage points. When the CME is low in the corona, the high cadence of images taken by SDO and STEREO nearly guarantees that simultaneous images are obtained from the three vantage points at regular five minute intervals starting from the flare onset at 01:25UT. \\

\indent  The dimensions of the ellipsoid are defined by a set of three parameters and its central position is defined in heliocentric coordinates (radius, latitude and longitude). An ellipsoid is considered a good visual fit when it intersects most of the red crosses. Off limb the ellipsoid must pass by the outermost extent of the CME. On disk the red crosses mark the location of the EUV front and must match the line of intersection of the ellipsoid with the solar surface. During the first 20 minutes of the event we used observations from STEREO and SDO.  Beyond the SDO AIA field of view (1.3 R$_\odot$), coronal images from Earth's perspective are obtained at low cadence by the LASCO coronagraphs. To cross-check the inferred location of the CME extent in LASCO images, we interpolated the four parameters at the LASCO C2 recording times; the interpolated locations are shown in the middle panel of row (f), revealing very good agreement between the observations and the fitted geometrical surface. In addition to the different time cadence of the different optical instruments, we noted that the signal to noise ratio in the COR-1 images is reduced near the edge of the field of view. This has been noticed before (e.g. Rouillard et al. 2010) and it can hamper our ability to accurately track the outer edge of the pressure wave in COR-1 when the CME reaches these heights. For this reason, we rely on COR-2 towards the external part of COR-1, where the COR-1 and COR-2 fields of view overlap as shown in the first column of rows (e) and (f). \\

\section{Overal comparison between shock location and the in-situ measurements:}

\begin{figure*}
\begin{center}
\includegraphics[angle=00,scale=.4]{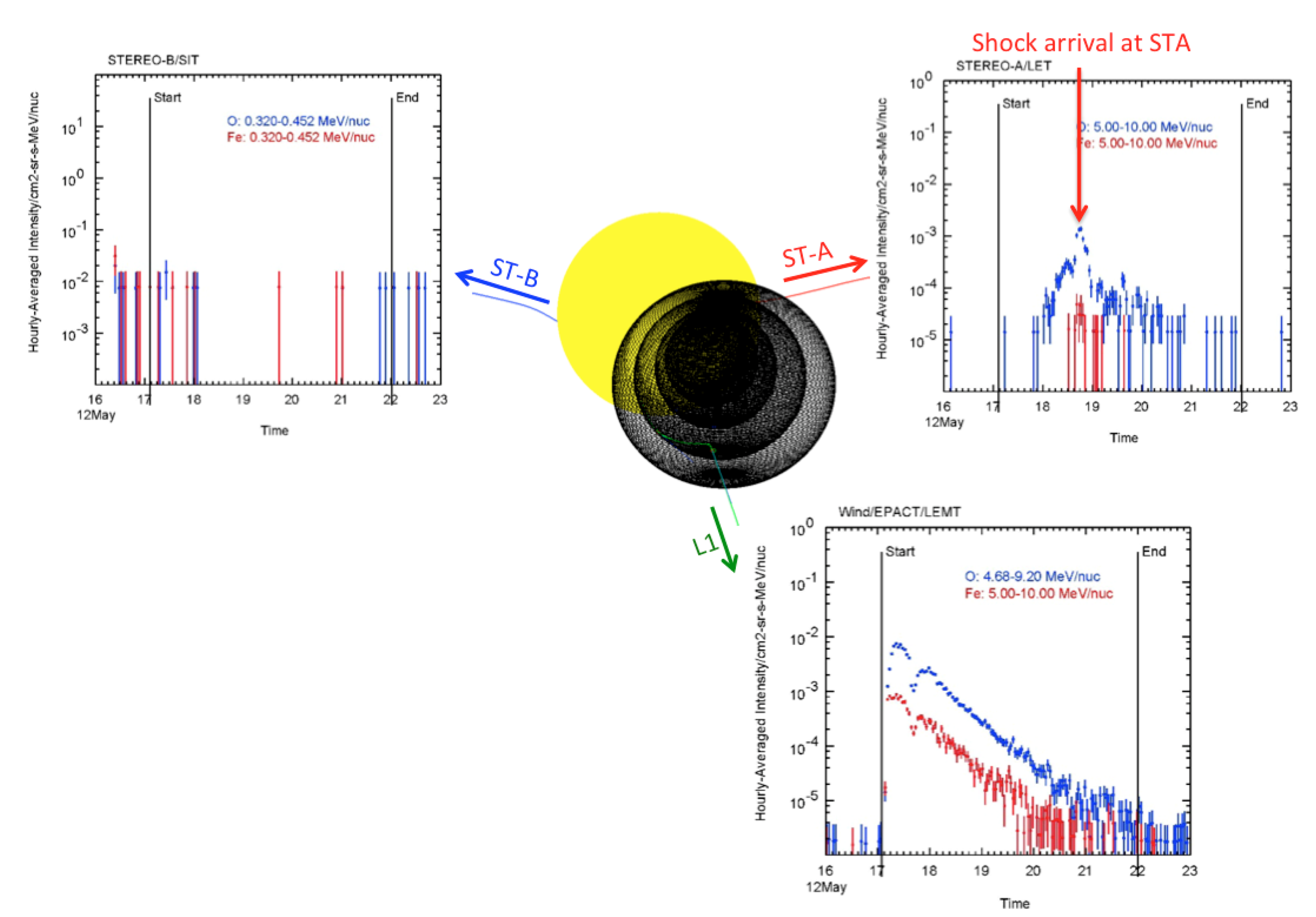}
\caption{Center: a view of the near-Sun environment with the triangulated locations of the propagating fronts at four successive times. The relative locations of the magnetic field lines connecting the STA, STB and L1 points assuming a  Parker spiral from the spacecraft to 2.5R$_\odot$ and the PFSS model from 2.5R$_\odot$ to the solar surface are shown as colored lines. The colored arrows mark the direction of hypothetical particles propagating outward towards the interplanetary medium.} Three panels show the time series of hourly-averaged 5-10 MeV/nuc Oxygen (blue lines) and Iron (red lines) fluxes measured over a 7-day interval by the LET instruments on the two STEREOs and the EPACT/LEMT instrument on the Wind spacecraft.
\label{SHOCKSEPs}
\end{center}
\end{figure*}

\indent Figure~\ref{SHOCKSEPs} presents, as superposed black ellipsoids, the location of the CME front at regular 5-minute intervals between 01:25 and 01:55 UT. We also show the Parker spiral connected with the ST-B (blue), ST-A (red) and near-Earth (green line) orbiting spacecraft. These spirals were defined by the speeds of the solar wind measured  in situ at the three spacecraft (ST-B: 300 km/s , ST-A: 350 km/s, and near Earth: 400 km/s) near the times shown in Figure~\ref{SHOCKSEPs}. Below 2.5 R$_\odot$, we trace the magnetic field lines using a Potential Field Source Surface (PFSS) model made available on solarsoft by the Lockheed Martin Solar And Astrophysics Laboratory (LMSAL)\footnote {$http://www.lmsal.com/~derosa/pfsspack/$}. The extrapolation is based on evolving surface magnetic maps into which are assimilated data from the Helioseismic and Magnetic Imager \emph{Helioseismic and Magnetic Imaging} (HMI; Scherrer et al. 2012) onboard SDO. These maps account for the transport and dispersal of magnetic flux across the photospheric surface  using a flux-transport model (Schrijver \& DeRosa 2003).The transport processes are differential rotation and supergranular diffusion, they modify continually the distribution of photospheric magnetic fields.  The area of the corona of interest in the present study is situated near the West limb, hence the photospheric magnetic field measurements used in the present study were only a few days old at the time of the extrapolation.

These estimated field lines allow us to determine approximately how the three spacecraft connect to the corona. According to Figure~\ref{SHOCKSEPs}, the  space environment situated near Earth is well connected with the emerging CME (green), whereas ST-A only connects with the CME much later and ST-B is not connected with the event. This is in qualitative agreement with particle measurements taken near Earth by EPACT/LEMT (ULEIS, Mason et al. 1998) on ACE and the Low Energy Telescope (LET, Mewaldt et al. 2008), one of four sensors that make up the Solar Energetic Particle (SEP) instrument of the IMPACT investigation on STEREO (Luhmann et al. 2008). The LET is designed to measure the elemental composition, energy spectra, angular distributions, and arrival times of H to Ni ions over the energy range from ~3 to ~30 MeV/nucleon.\\

The hourly-averaged flux of Oxygen ions in the 5-10 MeV/nuc energy range was very intense at the Wind spacecraft (10$^{-2}$ particles/cm$^2$-sr-s-MeV/nuc) , ST-A detected initially low Oxygen flux increasing steadily to peak at 2x10$^{-3}$ particles/cm$^2$-sr-s-MeV/nuc when the derived CME front intersects the spacecraft some 48 hours after the launch of the CME. In contrast ST-B measured no SEP event. Since this study is focused on the conditions that produced the GLE during the first few minutes of the CME launch, we do not discuss STA or STB particle measurements further, since the SEP either occurred much later for STA and not at all for STB.\\


\section{Properties of the emerging shock: the PFSS approach}
\label{sec:PFSS}
\subsection{Derivation of the 3-D shock speed}

\indent Once the parameters of the successive ellipsoids are obtained, we interpolate these parameters at steps of 150 seconds to generate a sequence of regularly time-spaced ellipsoids. To compute the 3-D expansion  speed of the surface of the pressure wave, we find for a point $P$ on the ellipsoid at time $t$, the location of the closest point on the ellipsoid at previous time-step $t-\delta t$ by searching for the minimal distance between point P and all points on the ellipsoid at time $t-\delta t$. We then compute the distance travelled between these two points that we divide by the time interval $\delta t=150$ seconds to obtain an estimate of the speed $P$. This approach slightly underestimates the shock speed calculated at time $t$ during the acceleration phase of the CME. We considered a set of 70$\times$70 grid points distributed over the ellipsoid. This number of points is computationally tractable and is sufficiently high to compute accurately the speed of the expanding shock as well as the shock geometry and Mach number described later. We compared this approach with that of computing the normal to the point $P$ at time $t-\delta t$ and finding the intersection between this normal and the ellipsoid at time $t$. Speeds computed by dividing the distance between this intersection at time $t-\delta t$ and $P$ at time $t$ gave nearly identical speeds to those computed with the minimal distance providing the heliocentric latitudinal/longitudinal coordinates of the center of the ellipsoid varies very slowly ($<5^\circ$) between consecutive 150 second time intervals, thereby guaranteeing that the consecutive ellipsoids are quasi-concentric. This condition is fulfilled for this event since we found that its latitude shifted southward from a heliocentric latitude of 4$^\circ$ at 01:30UT to stabilise at -2$^\circ$ at 01:45UT, while its longitude shifted eastwards from a Carrington longitude of 190$^\circ$ at 01:30UT to stabilize at 180$^\circ$ at 01:55UT.  \\

\begin{figure*}
\begin{center}
\includegraphics[angle=0,scale=.45]{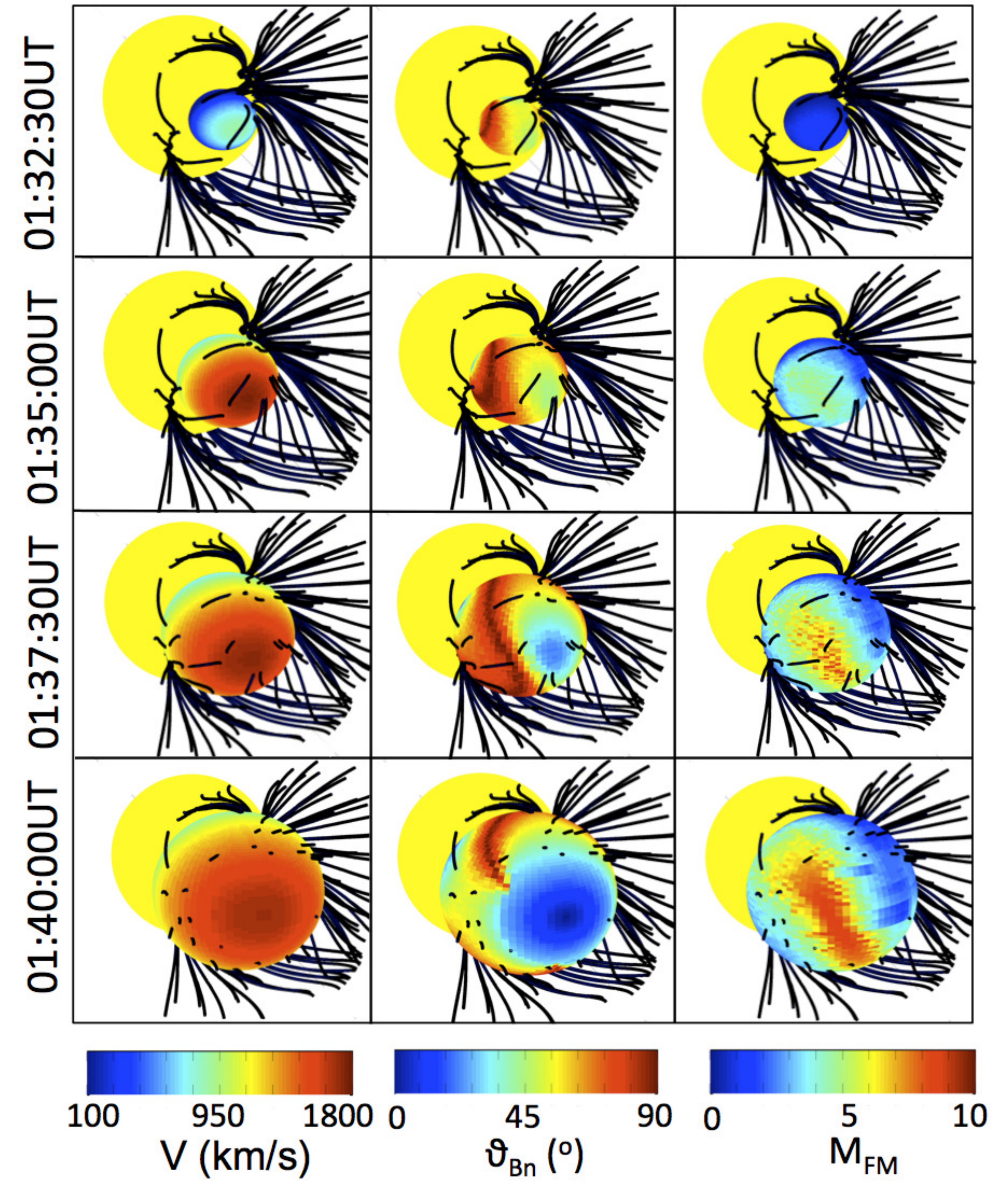}
\caption{The results of the derivation of shock parameters based on the combined inversion of imagery data and the PFSS model at four successive times during the eruption of the CME. Each column shows a different parameter: the shock normal speed (left), $\theta_{Bn}$ (center) and Mach number (right). Coronal magnetic field lines are traced in black.}
\label{MAthetaBnPFSS}
\end{center}
\end{figure*}

\indent Figure~\ref{MAthetaBnPFSS} presents the results of extracting the speed of the propagating front along the normal vector to the front surface and as a function of time. In addition to the location of the sphere, we trace open magnetic field lines using the PFSS model. Figure~\ref{MAthetaBnPFSS} shows that the CME front emerged from a region located inside a streamer.\\

\subsection{Derivation of the 3-D shock geometry and Mach number:}

We seek to derive the evolving properties of the CME-driven shock using the full set of available in-situ and imaging observations. In addition to the derivation of the shock speed, the parameters of interest are the shock geometry and the shock Mach number. The Mach number in an unmagnetised fluid is the ratio of the speed of the wave along the wave normal to the speed of sound of the ambient medium upstream of the wave. In a magnetised plasma, there are three  modes: the fast and slow magnetosonic waves and the intermediate Alfven wave. In this paper, the characteristic speed to which the front speed will be compared is the fast-mode speed, defined as:

\begin{equation}
V_{FM}=\sqrt{\frac{1}{2}\big[ V_A^2+C_S^2+\sqrt{(V_A^2+C_S^2)^2-4V_A^2C_S^2cos^2(\theta_{Bn})}\big ]}
\label{eq:Mach}
\end{equation}

where $V_A$ is the Alfven speed, $C_{S}$ is the sound speed, $\theta_{Bn}$ is the angle between the wave vector and the magnetic field vector. The Mach number, $M_{FM}$, is here defined as:

\begin{equation}
M_{FM}=\frac{V_S}{V_{FM}}
\end{equation}

where $V_S$ is the shock speed. We assume that at the very low coronal heights imaged here, the wind speed is zero. To derive the fast-mode speed, we need to derive the shock geometry, the properties of the background coronal plasma including temperature, density and the magnetic field. Since direct measurements of the 3-D coronal magnetic field strength are not yet possible, we have to employ some magnetic field reconstruction or modelling of the corona to infer the 3-D magnetic field distribution.\\ 

\indent A shock is quasi-parallel when $\theta_{Bn}<45^\circ$ and quasi-perpendicular when $\theta_{Bn}>45^\circ$. Equation 1 leads to the property that for a parallel geometry, the fast mode speed becomes $V_{FM}=\sqrt{ \frac{1}{2} \big[ V_A^2+C_S^2+\left|V_A^2-C_S^2 \right| \big]}$ whereas for a perpendicular geometry, $V_{FM}=\sqrt{V_A^2+C_S^2}$. For a coronal temperature of $T = 1.4$ MK, the sound speed is roughly 180 km/s, generally lower than the ambient Alfven speed except near the tip of streamers where the magnetic field strength can decrease by an order of magnitude. At this location, the shock becomes simultaneously quasi-parallel, in that region the fast mode speed is controlled by the sound speed. To derive the sound speed, we use $T = 1.4$ MK for the present PFSS approach. \\

\indent Coronal shocks undergo different regimes that are related to the value of their Mach number. There is a critical Mach number (M$_c$) (Edminston and Kennel 1984; Mann et al. 1995) above which simple resistivity cannot provide the total shock dissipation. The microphysical structure of collisionless shocks is very different when the shock is sub or super-critical (e.g. Marcowith et al. 2016). In the super-critical case a significant part of upstream ions are reflected on the shock front gaining an amount of energy that enables them to be injected into the acceleration process. Sub-critical shock do not reflect ions, significantly diminishing the ion and electron acceleration efficiency. M$_c$ is a function of the various shock parameters, but it has been argued that it is at most 2.7 and usually much closer to unity (e.g. Mann et al. 1995,  Schwartz, 1998). In the present study, a shock is said super-critical when $M_{FM}>3$. \\

\paragraph{Derivation of the ambient magnetic field properties:}

An extrapolation of the photospheric magnetic field to the corona using the PFSS technique can provide the magnetic field at all points on the surface of the triangulated shock surface. The PFSS model has a number of strong assumptions including a heliocentric spherical source surface, that no current is flowing in the corona and that the field is radial at the photospheric boundary (e.g. Wang and Sheeley, 1990). The line-of-sight component is measured by HMI and in the present study, the input is such that the line-of-sight component was converted to a radial component. It is common to also correct magnetograms for the poorly observed polar magnetic fields by applying a latitude dependent correction factor (Wang and Sheeley, 1991), here however no correction was applied since the magnetic maps used for the LMSAL PFSS model build up polar fields over time through transported processes. Finally, the measurements of surface magnetic fields are also prone to line profile saturation, including the HMI instrument, this saturation is not accounted for in the PFSS model used here. To derive the distribution of angles $\theta_{Bn}$ (Equation \ref{eq:Mach}) over the entire ellipsoid, we first derived numerically the vector normal to the ellipsoid at each point on its surface and then computed the angle between this normal vector and the coronal magnetic field vector obtained from PFSS at that point. \\

\indent The values of the magnetic field and plasma parameters modelled in the present section and in section \ref{sec:MHD} were interpolated between the numerical grid points at each location on the triangulated surface front. The modelled field is defined on a spherical grid $\{r_i,\theta_j,\phi_k\}$ with a constant step ${\rm d} \phi$ and variable steps in radial and co-latitude, i.e. ${\rm d} r$ and ${\rm d} \theta$ are non-constant. We adapted a linear interpolation method, described for a 2D axisymmetric grid by Cerutti et al. (2015), here generalized in 3D (volume weightening). The generalization is straightforward because the integration  over $\phi$ is elemantary. \\

\indent Given an arbitrary position $(r,\theta, \phi)$ where we want to interpolate the field value between grid positions, we find the cell in which the point is located, indexed as $\{i,j,k\}$ cell. We then calculate the total volume of the cell using:
\begin{eqnarray}
V_{\rm cell} &=&\int_{\phi_k}^{\phi_{k+1}} \int_{\theta_j}^{\theta_{j+1}} \int_{r_i}^{r_{i+1}} r^2 \sin \theta {\rm d}r  {\rm d} \theta {\rm d}\phi  \nonumber \\
            &=& {\phi_{k+1}-\phi_k \over 3} (\cos \theta_j-\cos \theta_{j+1})(r_{i+1}^3-r_i^3)
\end{eqnarray}

For the interpolation, 8 supplemantary sub-volumes are needed and are calculated in the same manner. The results are shown in the center column of figure \ref{MAthetaBnPFSS} and reveal that the geometry of the shock varies greatly in space and time. The shock is mostly quasi-perpendicular when situated below the streamers in closed field regions. It becomes quasi-parallel when the nose of the shock reaches the source surface near 01:37:30UT and enters open (and  (radial) field regions. A quasi-perpendicular geometry occurs mostly near its flanks as shown previously for other events (Kozarev et al. 2015). The band of high $M_{FM}$ is co-located with the region of quasi-perpendicular geometry but evolves within 10 minutes into a quasi-parallel geometry, we discuss this transition later in the paper. \\

\indent The Alfven speed is proportional to the ambient magnetic field strength and inversely dependent on the square root of the plasma density.  \\

\indent The open coronal fields computed with PFSS using uncorrected magnetograms can be at times much weaker than the open magnetic field measured near 1AU (Arden et al. 2014). Since we are interested in the process of shock formation along open magnetic field lines connected with near-Earth spacecraft, our approach has been to correct the magnetic fields derived from PFSS by using in-situ measurements. The total magnetic flux released into the interplanetary medium can be computed from PFSS extrapolations by simply averaging the unsigned radial field component at the source surface multiplied by its surface area. \\

\indent The expansion of the magnetic field leads to a more uniformly distributed radial field at the PFSS source surface than at the photosphere but the field has not yet spread out uniformly in latitudes and longitudes. The Ulysses spacecraft, that surveyed the radial component of the heliospheric magnetic field outside the ecliptic plane and as a function of heliospheric distance, revealed that beyond 1AU the absolute value of the radial field is independent of heliographic latitude (Smith and Balogh, 1995). This result implies that a re-distribution of the magnetic field continues beyond the source surface and for several tens of solar radii in the outer corona probably smoothing out differences in the tangential pressure and forcing the radial magnetic field to become uniform in latitude by 1AU. This redistribution occurs beyond the source surface and is more gradual than the strong re-distribution forced the source surface associated with the radial field boundary condition. As we shall see, MHD models with boundary conditions maintained at much higher coronal heights (30 R$_\odot$) suggest a more gradual redistribution of the radial field component with heliocentric distance than PFSS.\\

\indent The average of the radial field component extrapolated over the entire source surface is: 5.24 $10^{-6}$T at the time of extrapolation considered for this event. We also compared the full surface average (5.24 $10^{-6}$T) with an average of the radial field component taken over an area centered on the heliocentric coordinates of AR 11476 and extending 30 degrees around that region and we found an even lower value of 4.34 $10^{-6}$T or 82\% of the total surface average. We use this latter value to account for the possibility that the radial magnetic field may not have re-distributed uniformly over a sphere centered at the Sun and of radius 30 solar radii. \\

Since the Ulysses observations show that by 1AU the radial field measured in the ecliptic is representative of the radial field measured at any latitude, we can compare the average of the source surface field with the radial field values measured in situ in the ecliptic plane near 1AU. To derive the radial field value that is representative of the background magnetic field, we followed the procedure used in Rouillard et al. (2007) to derive the total open magnetic flux and averaged the absolute value of the hourly radial field values measured near 1AU over a full 27-day solar rotation period. The passage of large Interplanetary CMEs (ICMEs) will increase the background radial field values measured in situ near 1AU at a specific spacecraft. To obtain a more robust global estimate of the open magnetic field, we used not only the OMNI data but also the STA and STB magnetometer data. In all cases, the average radial field is close to 1.9$\pm$0.4 nT which we take as our reference radial field component representative of the 'background' solar wind  with a 20$\%$ uncertainty in this estimate. \\ 

\indent To compare the PFSS data to this radial field, we simply account for the nearly spherical expansion of the field between the source surface and 1AU such that the estimated radial field at 1AU is $ 4.34 \times10^{-6}\times(2.5/215)^2=0.58$nT, a factor of about 3.3 less than the measured radial field of 1.9nT near 1AU. A similar value is obtained by comparing the average open field at the reference source surface location (4.34 $\times 10^{-6}T$) in a region limited to the streamer where the CME originated with the value of the radial field measured near 1AU around the onset time of the SEP event. We conclude that the PFSS extrapolation used in the present study underestimates significantly the total open flux released in the interplanetary medium and that a correction factor of 3.3 should be applied to the PFSS data in order to obtain more realistic field values in the corona. The correction factor was obtained by comparing the radial components of the magnetic field at the source surface and at 1AU. To preserve the global topology of the field, the correction factor was applied to all components of the magnetic field including closed field regions of the corona that cannot be related to in-situ measurements made near 1AU. We adopted this technique because the focus of the present paper is on the production of SEPs that travel along open magnetic field lines to 1AU. In a future study, we will exploit radio imaging of other events to show that this correction may be too severe in the closed field regions of the corona. The correction will have the effect to substantially decrease the computed Mach numbers of the shock thereby providing conservative estimates. \\

\paragraph{Derivation of the ambient density:}

Past derivations of coronal densities have considered 1D  (Mann et al. 1999, LeBlanc et al. 1998) and 2D analytic models (Warmuth $\&$ Mann, 2005). These studies have shown that the use of a generic radial density model can lead to inaccurate derivations of local Alfven speeds due to the strong magnetic field gradients in the corona. In order to derive electron densities that are more representative of the (background) coronal conditions through which our pressure front is propagating we propose to invert remote-sensing observations. \\

\indent Estimates of the electron density distribution can be obtained by inverting EUV images using Differential Emission Measure (DEM) inverted from the SDO/AIA six coronal Fe filters (Aschwanden et al., 2001). For the density calculation using SOHO/LASCO, we use polarized brightness images. The brightness of the K-corona results from Thomson scattering of photospheric light by coronal electrons (Billings 1966). In the case of polarised brightness observations at small elongations (below 5 R$_\odot$ Mann 2003), the F-corona can be assumed unpolarized and thus does not contribute to the polarized signal; for this reason we restrict our derivation of electron densities to below 5 R$_\odot$. The technique employed to interpolate densities between the AIA and SOHO fields of view is detailed in the paper by Zucca et al. (2014). Beyond 5 solar radii, we assume that the plasma expands spherically to 1AU.  We assume that electrons situated within only 3 degrees longitude of the plane of the sky contribute to the emission.\\

\indent  In order to derive densities in the entire volume crossed by the triangulated front, we let the corona rotate in the plane of the sky for several days (spanning about 70 degrees of longitude) and repeat the aforementioned analysis every six hours (every 3.3 degrees of solar longitudes) between the 15 and 20 of May 2012. We then interpolate densities on a regular 1 degree longitudinal grid between each meridional plane to obtain a uniform 3-D grid of density values inside the entire volume crossed by the front. For this derivation, we checked that no large CME was present in the fields of view of AIA and C2 at the times used to derive the background densities. One of the assumptions made in this analysis is that the CME of interest here that passed through the LASCO field of view between 01 and 05UT did not alter the structure of the coronal streamer permanently and did not affect the density reconstruction between 17 and 20 May. This is supported by a smooth transition of the electron density variations derived before (00UT on 16 May) and after (06UT on the 17 of May) the CME passage. \\

\paragraph{The fast-magnetosonic Mach number:}

\indent The distribution of $M_{FM}$ is obtained by dividing the normal speed at each point on the triangulated front by the local fast mode speed of the medium. At these low coronal heights we can neglect the speeds of the solar wind plasma in this derivation of $M_{FM}$. The right-hand column of Figure~\ref{MAthetaBnPFSS} presents the distribution of $M_{FM}$ over the entire front. Before 01:32UT, the front is located well inside the streamer, the shock has not formed at these low heights ($M_{FM}<1$). Between 01:32 and 01:35UT, the front speed exceeds the local fast-mode speed in certain regions. This transition marks the formation of a shock and occurs near the onset time of the type II burst at 01:32UT (reference). A subset of the front reaches super-critical speeds ($M_{FM}>3$) when it enters the open magnetic field regions situated at the top of the streamer after 01:37UT. \\

\indent A band of high-$M_{FM}$ becomes very clear after that time along the front surface and located near the tip of the streamer and its associated neutral line. We found that the strongest rises in $M_{FM}$ mark drops in the background Alfven speeds. Such drops are mainly due to decreases in the strength of the coronal magnetic field and to a lesser extent to increases in the density since the Alfven speed is inversely proportional to the square root of the density.\\

\indent PFSS extrapolations show that the magnetic field lines that form the helmet streamers expand significantly between the photosphere and the source surface. We note that there is ample evidence that such a band of low magnetic field strength/high density (therefore enhanced plasma beta) exists from remote-sensing and in-situ measurements. The WL counterpart of this band is the plasma sheet typically observed as bright rays extending above helmet streamers (Bavassano et al. 1997; Wang 2009). The in-situ counterpart is thought to be the Heliospheric Plasma Sheet (HPS) typically measured during sector boundary crossings. This sheet is associated with very high plasma beta near 1AU due to an order of magnitude decrease in the magnetic field and signficant increases in the plasma density (Winterhalter et al. 1994, Crooker et al. 2004). The more the HPS/HCS is warped in latitude, usually in response to higher solar activity or weaker polar fields, the more the trajectory of a spacecraft making in situ measurements will be aligned to the normal of the local tangential plane of the HPS/HCS. During those times, the HPS measured near 1AU is well defined and of typically short duration, lasting at most 16 hours (Crooker et al. 2004). Assuming a typical rotation period of 25.38 days, we can convert that duration into a longitudinal width, this corresponds to about 10 degrees longitudinal width. The band of high $M_{FM}$ derived on the ellipsoid extends over a 10-15 degrees longitudinal width, near the upper limit of the size of the HPS typically measured near 1AU.  \\  

\indent The location of the source surface height at 2.5Rs is somewhat arbitrary. The justification for such a height stems from the coronagraphic observation that coronal electrons appear to flow rather radially beyond 2.5 R$_\odot$ (e.g. Wang and Sheeley 1990). Additionally, the position of coronal holes derived by PFSS agrees rather well with EUV observations and the sector boundary structure predicted by PFSS at 1AU agrees well with in-situ measurements during the different phases of the solar cycle (e.g. Wang et al. 2009). More recent studies have argued that a better agreement is obtained between the total open flux derived from the PFSS model and in-situ measurements by letting the source surface vary during the solar cycle (Arden et al. 2014). The large low-latitude coronal holes observed by STEREO during the solar minimum could be better interpreted by decreasing the height of the source surface to lower heights. Indeed decreasing the height of the source surface will allow more field lines to open to the interplanetary medium, this will increase the size of coronal holes and of the total open flux released to the interplanetary medium. \\

\indent We investigated the effect of changing the source surface height on the computed $M_{FM}$. The procedure described in section 4.1 to correct the total open flux values needs to be repeated for each new source surface height and we found correction factors ranging from 2.14 at 2Rs to 4.12 at 3Rs. We found that the band of high $M_{FM}$ retains its global shape for the three source surface radii, however lowering the radius to 2Rs induces a very broad (15-20$^\circ$) band of high $M_{FM}$. The broadness of this band of much lower magnetic field strength cannot be easily related with the heliospheric plasma sheet measured in the solar wind near 1AU. A heliospheric plama sheet measured for 16 hours and passing over a spacecraft at 300 km/s would correspond to a longitudinal extent of $10^\circ$. Increasing the source surface height to 3Rs delays the formation of the shock to after 01:32:30UT so that no shock has yet formed around the onset time of the type II bursts (01:32:00UT). A source surface height situated at 2.5 R$_\odot$ supports the existence of a shock already at 01:32:30UT and produces a broad but not too unrealistic band of high $M_{FM}$ perhaps akin to the heliospheric plasma sheet typically measured near 1AU. In addition we also checked that for a source surface at 2.5  R$_\odot$ the size of the coronal holes are similar to those observed by the EUVI instruments on STA.  \\

\section{Properties of the emerging shock: the MHD approach}
\label{sec:MHD}
\begin{figure*}
\begin{center}
\includegraphics[angle=0,scale=0.85]{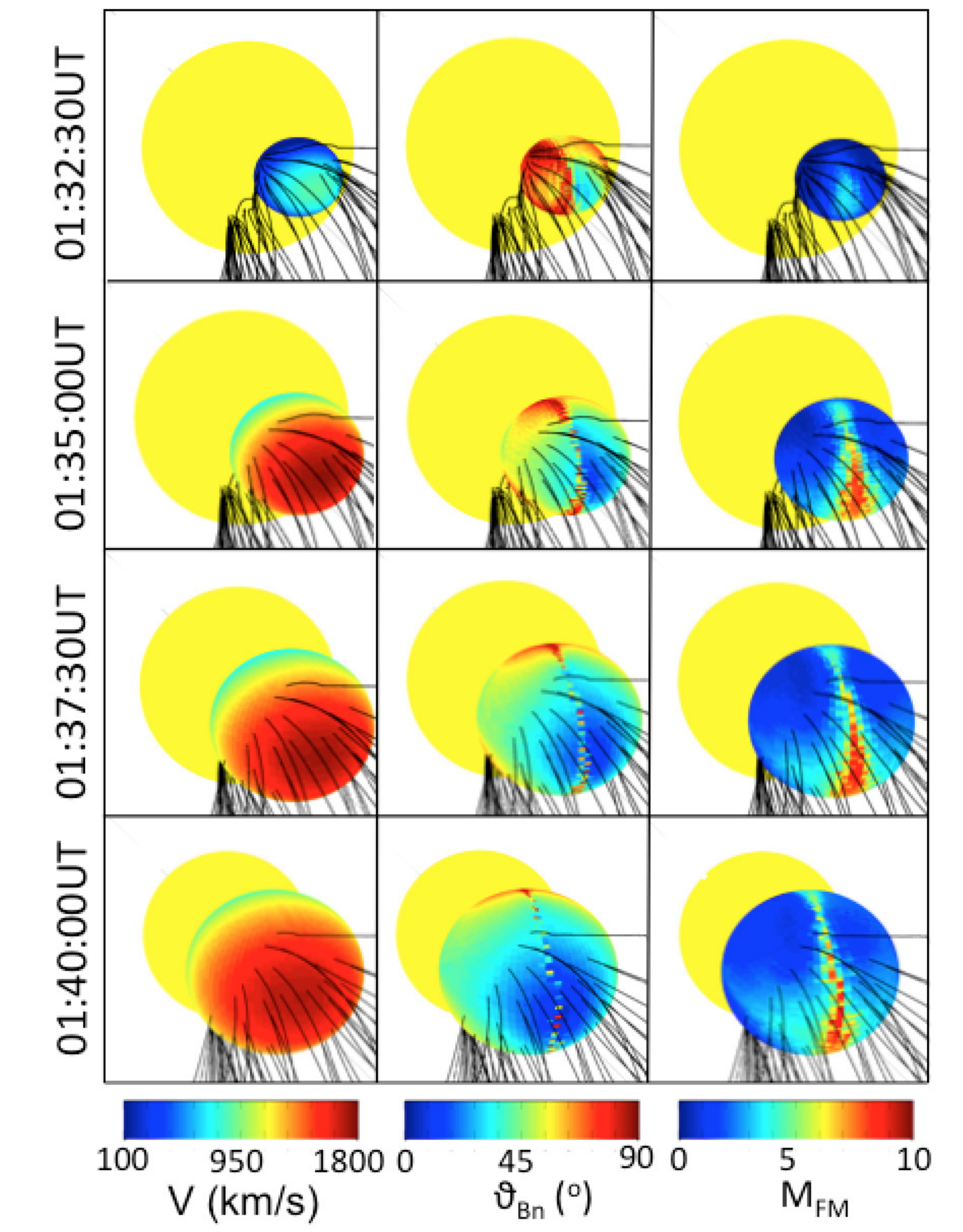}
\caption{In the same format as figure \ref{MAthetaBnPFSS} for the derivation of shock parameters based on the MAST MHD model.}
\label{MAthetaBnMHD}
\end{center}
\end{figure*}

A strong assumption of the PFSS model is the spherically uniform source surface that forces magnetic field lines to diverge rapidly from the photosphere to become radial at the surface. While coronagraphic imaging and in situ measurements provide strong supporting evidence for the existence of a narrow region of combined dense plasma and much weaker magnetic fields near the coronal/heliospheric neutral line, we investigated whether MHD simulations, with no source surface assumed, provide additional evidence for the formation of this region.  MHD simulations provide both derivations of the global magnetic field as well as plasma density and temperature.\\

\indent In this study, we used the two sets of 3-D MHD models developed by Predictive Sciences Inc. Like the PFSS model, these models use SDO HMI magnetograms as the inner boundary condition of the magnetic field. The outer boundary is set at 30 solar radii. The Magneto-Hydrodynamic Around a Sphere Polytropic (MASP) model is a polytropic MHD model and has a standard energy equation with a value of the polytropic index, gamma, close to 1 (typically 1.05) to crudely approximate the energy transport in the corona (Linker et al. 1999).  The temperature at the lower boundary in this model is selected to be a coronal temperature (1.8 mega-Kelvin). For the times of interest to this study the densities derived by this model tend to be unrealistic. Indeed, applying an inverse square density fall off between 30 R$_\odot$ and 1AU to compare the model with in-situ measurements, we find that simulated densities are an order of magnitude too high compared with those measured in the solar wind. \\

\indent The Magneto-Hydrodynamic Around a Sphere Thermodynamic MAST model is a MHD model with improved thermodynamics including realistic energy equations with thermal conduction parallel to the magnetic field, radiative losses, and coronal heating. The effect of Alfven waves on the expanding coronal plasma is also included using the so-called Wentzel-Kramers-Brillouin approximation. The temperature at the lower boundary in this model is 20,000 K (approximately the upper chromosphere), and the transition region is captured in the model.  Special techniques are used to broaden the transition region such that it is resolvable on 3D meshes and still gives accurate results for the coronal part of the solutions. The coronal heating description is empirical and the coronal densities arise entirely from the heating and its interaction with the other terms.  A description of this model appears in Lionello et al. (2009). \\ 

\indent Extrapolating the simulated values from the outer boundary of the model (30Rs) to 1AU reveals that, like for PFSS, the simulated neutral line maps to the heliospheric current sheet. In addition, the density values are also well simulated and fall in the range of density values measured before the onset time of the SEP event. The average value of the coronal magnetic field threading a sphere centered at the Sun and located at 5 R$_\odot$ leads to values of $\sim$1.3nT at 1AU. We choose a height of 5 R$_\odot$  after tracing open and closed magnetic field lines in the MAST model; we found that beyond this height magnetic field lines are mostly open to the interplanetary medium. We remind the reader, that in contrast to MHD models, the height at which magnetic field lines are all open is set by that of the source surface in the PFSS model. The average radial field measured in MAST is lower than the measured radial field values near 1AU (1.9$\pm$0.4 nT). A correction was applied to magnetic field values of the MAST model by multiplying all field values by a factor of 1.5($=1.9/1.2$). Again the correction factor is applied to all components of the magnetic field to preserve the global topology. \\

\indent  We show in Figure \ref{MAthetaBnMHD}, the front speed (left),  $\theta_{Bn}$ (center) and $M_{FM}$ (right) on the surface. In the MHD model the neutral line forms at the same location as in the PFSS model, but is more oriented along the North-South direction than the neutral line derived with PFSS. Just like for the previous technique, the fast-mode speed drops to low values in the vicinity of the neutral line ($<$200 km/s), thereby boosting $M_{FM}$ because the magnetic field strength is low and the density high. We note that the field strength drops in this region due to a combination of the field expansion (like PFSS) and, since we are using a MHD model, some level of numerical diffusion which forces field lines to reconnect. Although here not physically resolved, such reconnection processes are thought to occur in the vicinity of the real neutral line and of the heliospheric plasma sheet since complex magnetic structures reminiscent of magnetic flux ropes and field line disconnections are frequently measured near 1AU in the heliospheric plasma sheet (Crooker et al. 1996, Rouillard et al. 2011c). The MHD model suggests additionally that the Alfven speed drops to 200-300 km/s along the southern flank of the CME structure. The $M_{FM}$ values typically range from below 1 to beyond 7 across the triangulated 3-D front after 01:37:30UT with the highest values occurring in the vicinity of the neutral line. This is in general agreement with the PFSS/DEM technique presented in the previous section.  \\

\indent Finally, comparing the middle and right-hand columns of Figure \ref{MAthetaBnMHD} shows that the highest $M_{FM}$ tend to occur for a quasi-parallel geometry. We note however a region of oblique to quasi-perpendicular shock and high-Mach number along the southern flank of the structure. This contrasts with PFSS that predicted a similar band of super-critical quasi-perpendicular shock across the nothern flank of the CME. \\

\section{Comparison between the emerging shock and radio measurements:}

\begin{figure*}
\begin{center}
\includegraphics[angle=0,scale=.55]{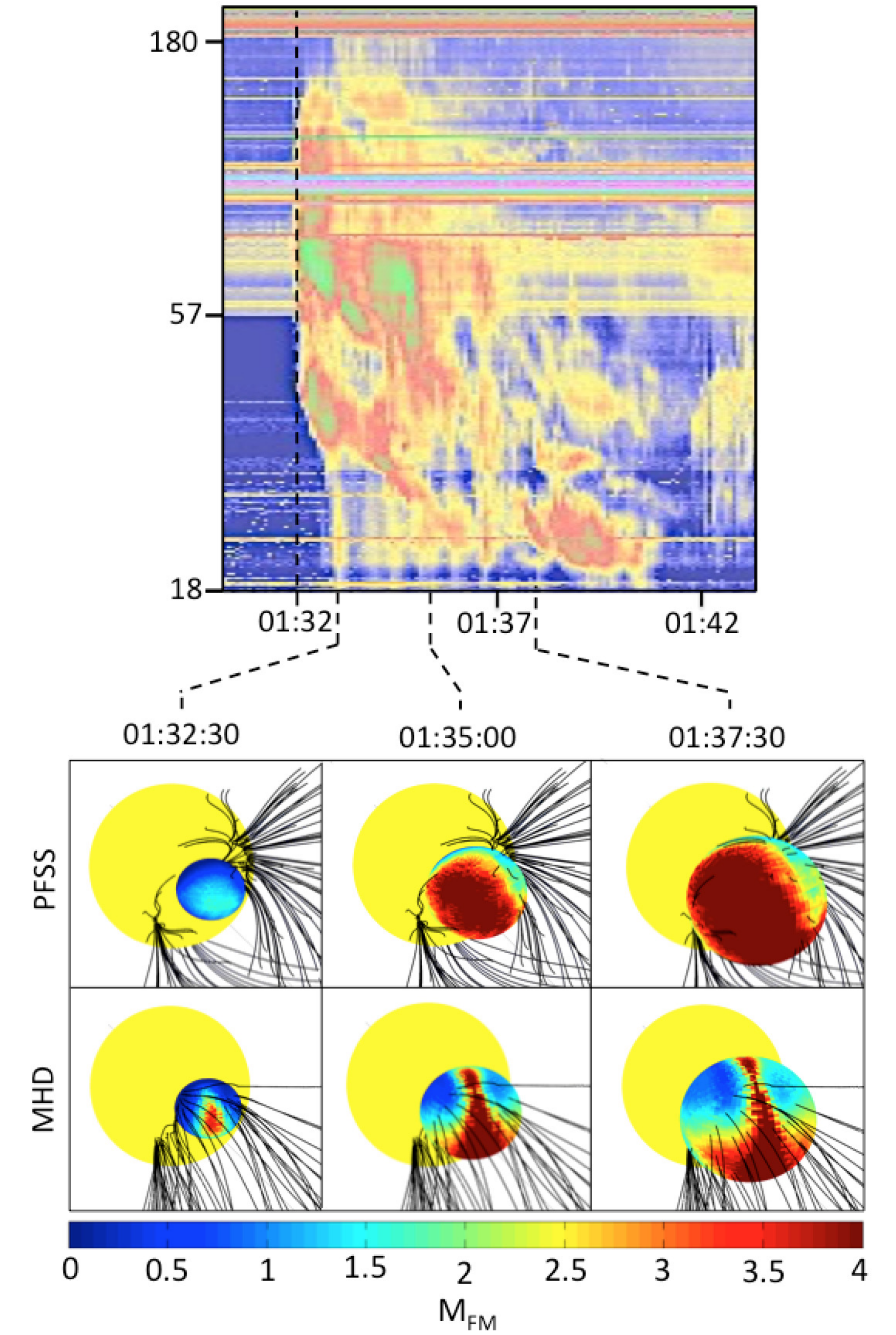}
\caption{Top: a radio spectrogram from the Culgoora radioheliograph showing the type II burst starting at 01:32UT. Bottom: the distribution of M$_{FM}$ over the propagating front at the three first times (01:32:30UT, 01:35UT, 01:37:30UT) shown in the right-hand column of Figures \ref{MAthetaBnPFSS} and \ref{MAthetaBnMHD}. The color table was defined on a smaller range of values of M$_{FM}$ between 0 and 4 for this early phase of the eruption. Magnetic field lines are shown in black.  }
\label{typeIIMA}
\end{center}
\end{figure*}

As mentioned in the introduction and visible in the spectrogram of Figure \ref{typeIIMA}, type II bursts were measured during the event by the Culgoora radioheliograph starting at 01:32UT and drifting from 140 to 18 MHZ. Both the fundamental and the harmonic are visible on the spectrogram. For comparison, the estimated distribution of $M_{FM}$ on the shock surface are repeated from Figure \ref{MAthetaBnPFSS} and  \ref{MAthetaBnMHD} for the two models used, but we changed the range of the color table from  $M_{FM}=0$ to  $M_{FM}=4$. For the technique used here, the earliest time at which Mach numbers were derived was 01:32:30UT so 30 seconds after the onset of the type II burst and both models confirm that a part of the pressure wave has become a shock, with a maximum of $M_{FM}\sim1.5$ for PFSS and some more localised increases of $M_{FM}\sim3$ for MHD. The shock is sub-critical for the PFSS technique but is already becoming super-critical in a limited area near the nose of the pressure wave in the MHD approach. As the event evolves the shock becomes rapidly super-critical over large fractions of the surface in both approaches. Multiple portions of the pressure wave are becoming super-critical, this is perhaps providing an explanation for the complex nature of the type II at 01:32UT observed in this spectrogram. We also investigated how far our technique could explain the sudden onset of the type II burst by using additional SDO AIA images with 30 second time cadence at intermediate times between 01:30:00UT and 01:32:30UT but without SECCHI data available at such high cadence, the results of this analysis were not sufficiently conclusive as they were too limited by the single viewpoint. In essence the details of the initial expansion rate were not sufficiently resolved at these intermediate times.

\section{Deriving shock parameters along specific field lines:}

\begin{figure*}
\begin{center}
\includegraphics[angle=0,scale=.40]{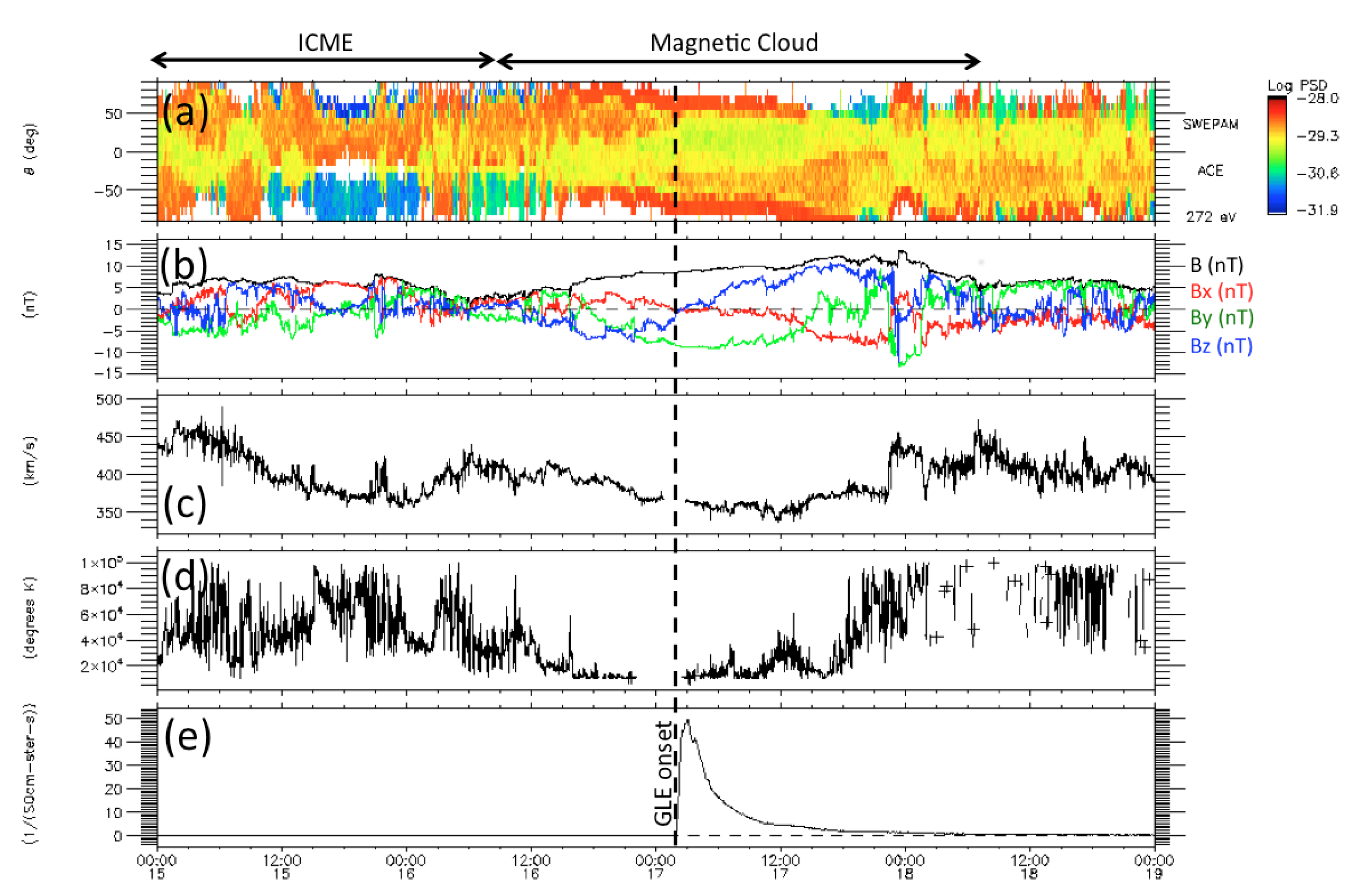}
\caption{Properties of the interplanetary medium measured near Earth over a 4 day-interval centered on the onset of the GLE event. The parameters shown are the normalised spectrogram of suprathermal electrons measured by ACE SWEPAM at 272 eV (a), the magnetic field magnitude (black) and components (b) measured by Wind MFI, the plasma speed measured by Wind (c), the temperature measured by Wind (d) and the flux of particles with energies exceeding 60 MeV  measured by GOES (d).}
\label{INSITUSEP}
\end{center}
\end{figure*}

\indent To compare the properties of energetic particles measured near 1AU with the properties of shocks inferred in the corona, we need to consider the path followed by energetic particles to propagate from the Sun to 1AU. Since these escaping particles gyrate along interplanetary magnetic field lines, a model for the interplanetary magnetic field is usually assumed; the simplest and most common approach is to model these field lines as an Archimedian spiral. The locus of these spirals is controlled by two parameters, the speed of the solar wind carrying the field line of interest and the rotation rate of the Sun. The speed is usually defined by the average solar wind speed measured in situ at the onset time of the SEP event. Typically the spiral connects at outer boundary of the coronal model used (2.5 R$_\odot$ for the PFSS model and 30 R$_\odot$ for the MHD model) and the spacecraft making the in-situ measurement.\\

\indent The assumption of an Archimedian spiral to connect near-Earth data with the shock requires that the SEP event occured during quiet solar wind conditions, both in the near-Earth environment and in the region situated between the Sun and Earth. However solar wind measurements made in situ near Earth reveal that a magnetic cloud was passing at the time of the GLE onset (Figure \ref{INSITUSEP}). The ACE spacecraft measured several common signatures of magnetic cloud including counter-streaming electrons (Figure \ref{INSITUSEP}a), a smooth rotation of the magnetic field (Figure \ref{INSITUSEP}b,c,d) and a low temperature (Figure \ref{INSITUSEP}e). The period preceding that magnetic cloud passage may also be another Interplanetary CME passage since complex magnetic fields and atypical suprathermal electron signatures were also measured at the time. \\

\indent We investigated the origin of these transients and whether they also erupted in AR 11476 that later produced the 17 May 2012 and its GLE event. The aim being to determine if magnetic connectivity between the vicinity of AR 11476 and the ACE spacecraft were likely to be established by the internal field of the magnetic cloud measured during the SEP event. The heliospheric imagers onboard STEREO were imaging this region continuously days before the GLE event and allow CMEs and CIRs to be located in 3D (e.g. Rouillard et al. 2008; 2011a,b). We considered the CME and CIR catalogues made available by the Heliospheric Cataloguing, Analysis and Techniques Service (HELCATS) FP7 project. This project produced the first systematic catalogues of CMEs  (Harrison et al. 2016)  and CIRs (Plotnikov et al. 2016) observed by the heliospheric imagers onboard STEREO. A detailed analysis of the state of the interplanetary medium days preceding and during the GLE event is presented in Appendix B for clarity purposes, since this paper focuses mostly on the 3-D expansion of the shock and its effect on energetic particles. The conclusion of the analysis is that, during the SEP event, the near-Earth environment is magnetically connected to the region where the shock forms by a magnetic cloud that erupted five days earlier (on 12 May 2012) from the vicinity of the same AR 11476 (see Appendix A for more details).  \\

\begin{figure}
\begin{center}
\includegraphics[angle=0,scale=.536]{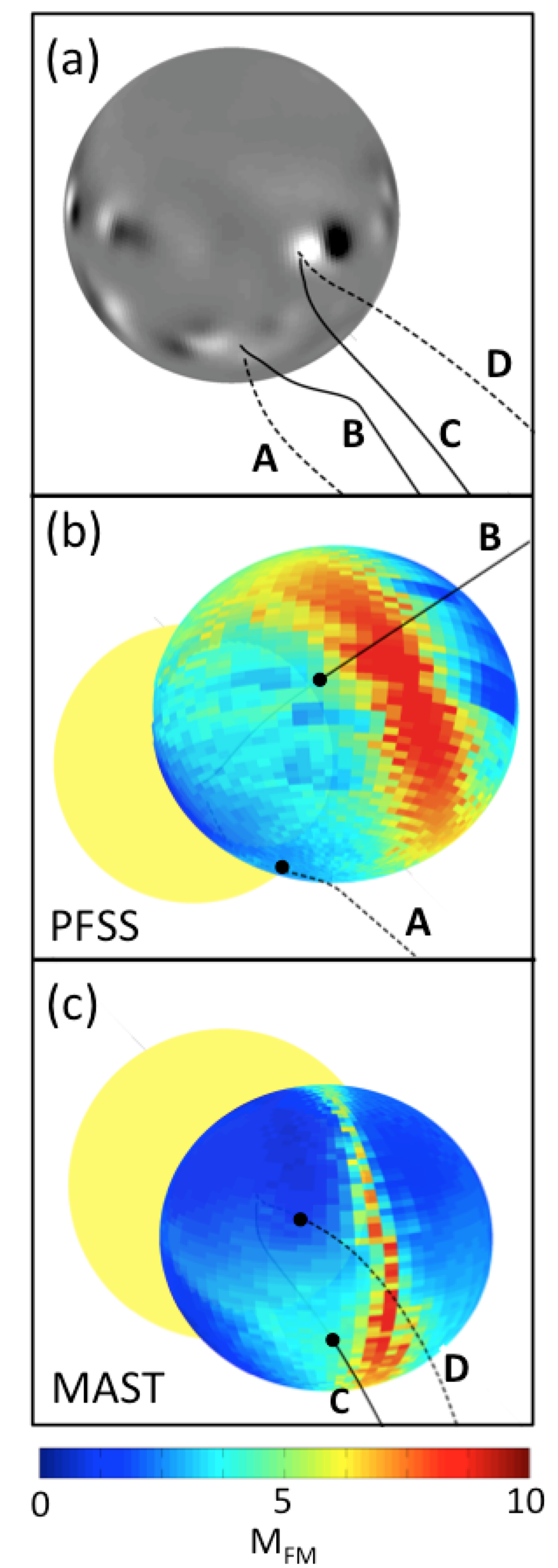}
\caption{A smoothed HMI magnetograms projected on the solar disk with the position of the shock triangulated in this study at 01:45UT. Two magnetic field lines are derived from the PFSS (left) and the MAST MHD model (right). For each image, the black line passes through the band of high-Mach number while the dashed line passes through a weaker part of the shock. }
\label{FLshock}
\end{center}
\end{figure}
\indent The passage of a magnetic cloud at GLE onset makes a tracing of the field line linking the Earth to the low corona impossible with our current limited understanding of the internal structure of CMEs. Instead we decide to illustrate the variability of shock properties along different open magnetic field lines, by extracting shock parameters along two different lines for each model. These four lines are traced in Figure \ref{FLshock}a with a smoothed HMI magnetogram shown on the surface of the Sun, the strong bipole (black/white region) is AR 11476. Magnetic field lines (A,B) and (C,D) are open to the interplanetary medium and are from the PFSS and MHD models respectively. For both models, the solid lines pass through the region of high-Mach number while the dashed lines pass through low-Mach number.  In addition to these magnetic field lines, Figures \ref{FLshock}b and \ref{FLshock}c show the reconstructed $M_{FM}$ values from, respectively, the PFSS (Figure \ref{MAthetaBnPFSS}) and the MHD (Figure \ref{MAthetaBnMHD}) models.   \\

\begin{figure*}
\begin{center}
\includegraphics[angle=0,scale=.52]{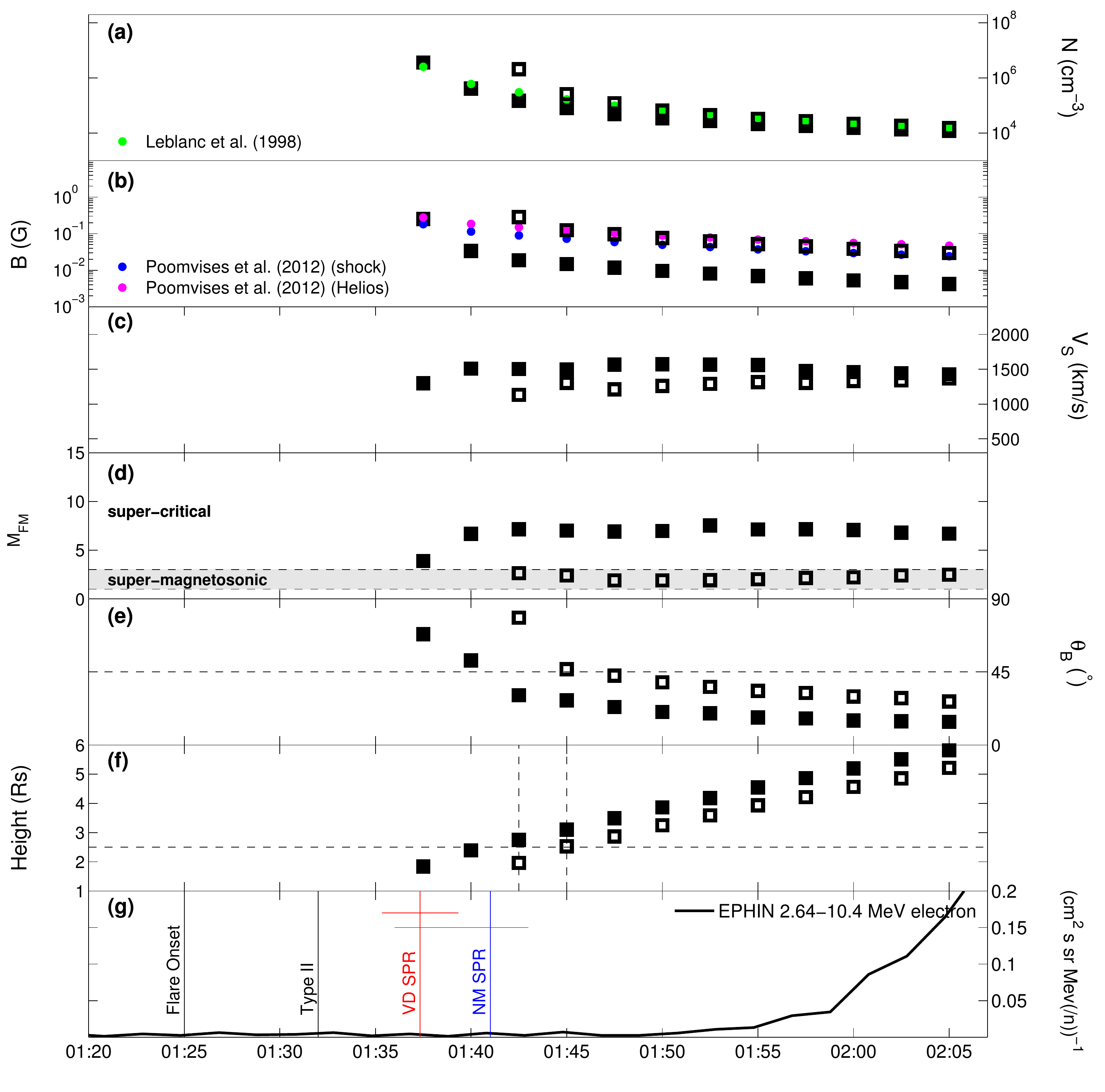}
\caption{Shock plasma parameters extracted using the PFSS/DEM technique at the intersection between the triangulated shock and a magnetic field line passing through the band of high Mach number shown in Figure\ref{MAthetaBnPFSS} (filled squares) and  a magnetic field line intersecting the shock far from that band (open squares). These parameters are plotted as a function of time (in UT). Panels (a) and (b): the ambient coronal density (N, cm$^{-3}$) and magnetic field (B, G) upstream of the shock. Superposed on these plots are derivations of ambient plasma properties from other studies as detailed in the text. Panels (c), (d), (e) and (f) show the shock speed (V$_s$, kms$^{-1}$), Mach number ($M_{FM}$), $\theta_{Bn}$, heliocentric distance (R$_{\odot}$)   at the intersection between the shock and different magnetic field lines of the helmet streamer. Panel (g): the flux of 2.64-10.4 MeV electrons as a function of time with superposed the times of the flare onset, and Type II burst. The SPR times derived by Gopalswamy et al. (2013) (GEA SPR) and derived by the velocity dispersion analysis in Appendix A (VD SPR) are shown as vertical blue and red lines, respectively. The uncertainty in these estimates are shown as the corresponding horizontal segments. }
\label{PFSSFLC} 
\end{center}
\end{figure*}
\begin{figure*}
\begin{center}
\includegraphics[angle=0,scale=.52]{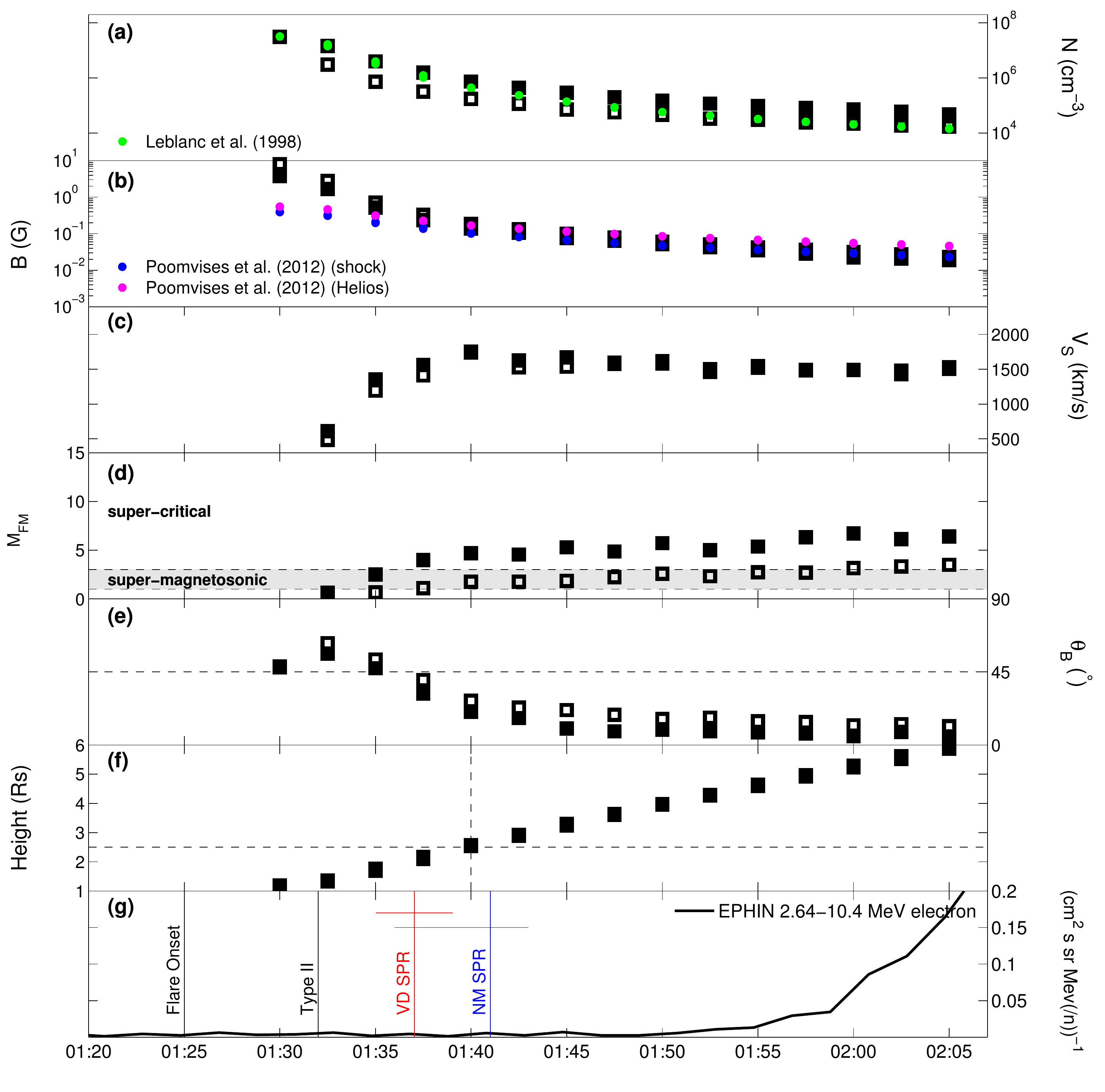}
\caption{In exactly the same format as Figure \ref{PFSSFLC} but for the parameters extracted using the MAST MHD model from Predictive Sciences Inc. Just as in Figure \ref{PFSSFLC}, the filled squares correspond to a field line passing inside the band of high-Mach number shown in Figure\ref{MAthetaBnPFSS} while the open squares correspond to a field line passing the shock far from the band.}
\label{PSIFLC}
\end{center}
\end{figure*}

\indent The open and filled squares in Figure \ref{PFSSFLC}  and \ref{PSIFLC} correspond to shock parameters extracted along the dashed and continuous field lines in Figure \ref{FLshock}. Figure \ref{PFSSFLC}a shows the background coronal density at the shock-field line intersection derived using the differential  emission measure. For comparison, the green dots show the densities that are obtained when assuming the Leblanc et al. (1998) profile at the height of field line-shock intersection (Figure \ref{PFSSFLC}f). The Leblanc et al. (1998) profile was derived from the drift of type III bursts and assumes a density at 1AU of about 7 cm$^{-3}$, very close to the density measured near 1AU. Our two curves of reconstructed densities differ initially by an order of magnitude but they rapidly converge  with the Leblanc et al.'s densities above 3Rs.\\

\indent Figure \ref{PFSSFLC}b shows the background coronal magnetic field at the shock-field line intersection using the PFSS model. For comparison, the purple and blue diamonds show the magnetic fields derived from the relation of Poomvises et al. (2012) at the height of field line-shock intersection (Figure \ref{PFSSFLC}f). The Poomvises et al. (2012) profiles were derived from the stand-off distance between CME core and the driven shocks. The reader is referred to their paper for more information. We find that the field line threading the shock far from the neutral line is very similar to the Poomvises et al. (2012) profile while the magnetic field strength of the line passing near the neutral line is systematically an order of magnitude lower. Figure \ref{PFSSFLC}c shows the variation of the shock speed along the two field lines, due to the rather spherical expansion of the shock, the two speeds do not differ much between the field line locations. $M_{FM}$ is strongly dependent on the magnetic field strength. The very different magnetic field strengths are therefore reflected in the $M_{FM}$ values that are much higher for the field line passing near the neutral line.  \\

\indent The flux of relativistic electrons (2.64-10.4 MeV) shown in Figure \ref{PFSSFLC}g were obtained by the Electron Proton Helium Instrument (EPHIN) part of the \emph{Suprathermal and Energetic Particle Analyzer} (COSTEP) (M\"{u}ller-Mellin et al. 1995 onboard SOHO ({PFSSFLC}d). In addition, the onset times of the flare and type II burst and the estimated Solar Release Time (SPR) of the GeV protons are is shown in Figure \ref{PSIFLC}d). According to PFSS the shock is initially confined to closed-field regions but at roughly 01:37 the shock enters the open field regions. At the time, the geometry is quasi-perpendicular (Figure \ref{PFSSFLC}c), but $M_{FM}$ is still small. It is not until 01:37:30UT, about the time of the SPR time, that the Mach number increases dramatically along field line connected to the vicinity of the neutral line. Also shown in Figure \ref{PFSSFLC}g) are the SPR times derived by Gopalswamy et al. 01:41 (+00:02/-00:05) UT and in Appendix A, 01:37:20 (+00:02/-00:02) UT, using the velocity dispersion analysis. The SPR times occur during the transition to super-criticality particularly along the field line passing near the neutral line. In addition the analysis suggests that the shock was quasi-perpendicular at the estimated SPR.\\

\indent Figure \ref{PSIFLC} is in an identical format to Figure \ref{PFSSFLC}  but it displays the results of extracting shock parameters using the MHD simulation instead of the PFSS model. The field line passing in the vicinity of the neutral line shows higher density values than the Leblanc et al. (1998) profile, however the density values are lower for the field line far from the line. Like PFSS the magnetic field magnitude is generally lower for the field line passing near the neutral but the difference is less pronounced between the two field lines than for PFSS. As already seen in Figure \ref{FLshock}, the connectivity between the triangulated front and open field lines is established much earlier than for PFSS at 01:30UT or about 5 minutes after the flare onset. At that time, the geometry is quasi-perpendicular (Figure \ref{PSIFLC}e) but the front is sub-critical (Figure \ref{PSIFLC}d). At the formation time of the shock,  $\theta_{Bn}$ is close to 45$^\circ$ which points to the occurrence of an oblique shock. The variation of $M_{FM}$ is more gradual, the shock becomes super-critical for both field lines considered near the SPR release time of GeV particles. The SPR times occur for this model when the shock becomes super-critical just after the quasi-perpendicular phase when the shock has reached an oblique geometry.

\section{Discussion:}

The images taken by the STEREO and near-Earth orbiting spacecraft are sufficient to map the 3-D extent of propagating fronts that form during the eruption of CMEs. According to our geometrical fitting technique, the shape of the propagating front remained highly spherical during the eruption process. This is also seen in the event analysed by Kwon et al. (2015). In our analysis, the EUV front is considered to be the low-coronal signatures of the expanding WL front (see Figure~\ref{TRIANGWL}). During the 17 May 2012 event, the speed of the EUV front never exceeded 500 km/s (Figure 5) and the fastest lateral motions are not measured in EUV images but higher up in the corona  (Figure \ref{MAthetaBnPFSS}). \\ 

\indent The novelty of the present technique, also exploited in Salas-Matamoros et al. (2016), resides in the derivation of the normal speed and the Mach number ($M_{FM}$) over the entire surface of the CME front (Figure \ref{MAthetaBnPFSS}  and \ref{MAthetaBnMHD} ). This was obtained from a combination of different techniques incuding the inversion of coronal imaging, magnetic field reconstructions and MHD modelling. The CME drives a shock and even a super-critical shock with $M_{FM}$ values in excess of 3 (Figure \ref{MAthetaBnPFSS}  and \ref{MAthetaBnMHD} ) with the highest values occurring near the nose of the CME. This is in agreement with the results of Bemporad et al. (2014) who used remote-sensing observations from SOHO to derive the coronal and shock properties of the 11 June 1999 CME in the plane of the sky. Our analysis shows additional structure in the distribution of the $M_{FM}$ induced by the complex topology of the background field. At the earliest time available (01:32:30UT), portions of the triangulated front have already steepened into a shock, this is in agreement with the detection at the time of a type II burst (Figure \ref{typeIIMA}). \\  

\indent It is interesting to compare this result with the analysis of the 2011 March 21 CME event (Rouillard et al. 2012). In that analysis, the expanding front was not fitted using the technique presented in this paper, however the shape of the CME, as seen projected in the plane of the sky, appeared highly elliptical with its major axis orthogonal to the direction of propagation (see Figure 2 of Rouillard et al. 2012). The very strong lateral expansion of the front observed in WL tracks also the EUV front. This lateral expansion eventually pushed streamers located far from the source region, at the same time that the EUV wave reached the footpoints of those streamers. That lateral expansion speed was on average about 400 km/s, but the pushed streamers were launched with a speed of 900 km/s. That paper demonstrated that the speed parallel to the solar surface of the EUV and WL  fronts was at the same. However contrary to the EUV wave that moves only along the solar surface, the pushed streamers have an additional high radial speed component. For the 17 May 2012 event, analysis of the Mach numbers (not shown here) at the very low height imaged by EUV instruments remains predominantly with $M_{FM}<1$ throughout the event with small patches of 1$<M_{FM}<$2, therefore while certain parts of the EUV wave may have steepened into a shock, it remains sub-critical throughout the event. The analysis presented in the present paper suggests that a super-critical shock is unlikely to develop at the heights we observe EUV waves but that a shock can rapidly become supercritical at the heights imaged by WL coronagraphs (2R$_\odot$) near the tip of streamers.\\

The formation of a super-critical shock means that early during the eruption process, instabilities  develop along the shock front that could play a role in the acceleration of high-energy particles. Preliminary simulations that model the process of diffusive-shock acceleration (Sandroos and Vainio, 2009) using the magneto-plasma properties of the shock derived in the present paper suggest acceleration to 300 MeV in 80 seconds on some of the open magnetic field lines. \\

\indent For both models, the rapid rise of $M_{FM}$ values occurs when the propagating front reaches open magnetic field that diverge strongly near 2-3Rs where helmet streamers typically form. The highest values of $M_{FM}$ are associated with the coronal neutral line, the source location of the heliospheric current sheet and its surrounding heliospheric plasma sheet. The latter, frequently measured near 1AU, is typically associated with magnetic fields that are an order of magnitude smaller than those measured in the ambient solar wind. This is clearly predicted by the PFSS reconstruction. However the PFSS model may over-estimate the size of this region. Both PFSS and the MHD approach reveal that the $M_{FM}$ values extracted along open field lines crossing the shock far from the neutral line remain overall below super-critical values ($<$3) until high up in the corona ($>$2R$_\odot$).\\

\indent Past studies have used SOHO observations to infer the heights of WL CMEs at the onsets of GLEs measured since 1997 (Gopalswamy et al. 2012). These inferred heights were obtained from a single viewpoint and therefore less accurate than the technique used in the present paper. Additionally, the limited coverage of the corona obtained from SOHO required interpolation techniques. Nevertheless these approaches provide estimates of the time delay between the onset of the flare and the onset of the GLE. These heights are listed in Table 1, columns 2 and 3 list the height estimates made by  Gopalswamy et al. (2012) and Reames et al. (2009), respectively, using different approaches. They found the heights of particle releases above 2-3Rs showing a long delay between the onset of the flare and the injection time of high-energy particles. The results of the present paper provides evidence for the delayed release times of GeV protons to be related with the time needed for the shock to become super-critical.\\

\begin{deluxetable*}{ccrrrrrrrrcrl}
\tabletypesize{\scriptsize}
\tablecaption{Characteristics of WL CMEs and in situ measurements during GLE events}
\tablewidth{0pt}
\tablehead{
\colhead{GLE no } & \colhead{ Date} & \colhead{Time} & \colhead{CME Ht at SPR} & \colhead{CME Ht at SPR} &
\colhead{State}  }
\startdata
55 &1997 Nov  06 	& 	110000  -  120000	& NG  			&	2.34 		&	Solar Wind, before
ICME 	& 8 	 & No\\
(1) &(2) 	& 	(3)	& (4)  			&	(5) 		&	(6) 	 \\
\hline
56 &1998 May  02 	& 	110000  -  120000 	& 2.9 $\pm$ 0.2  	&	1.97 		&	Inside ICME  \\		
57 &1998 May  06 	& 	070000  -  080000   	& 2.0 $\pm$ 0.2  	&	2.21 		&	Inside ICME \\
58 &1998 Aug   24 	&	190000  -  200000   	& 5.7 $\pm$ 0.5  	&	5.14 		&	Solar Wind \\
59 & 2000 Jul   14 	& 	090000  -  100000  	& 2.6 $\pm$ 0.3  	&	1.74 		&	Inside ICME\\
60 & 2001 Apr  15 	& 	110000  -  120000   	& 2.4 $\pm$ 0.2  	&	2.10 		&	Inside ICME\\
61 & 2001 Apr  18 	& 	010000  -  020000   	& 4.8 $\pm$ 0.7  	&	3.92 		&	Inside ICME\\
62 & 2001 Nov 04 	& 	150000  -  160000   	& NG  			&	8.05 		&	Between two ICMEs \\
63 & 2001 Dec 26 	&  	040000  -  050000   	& 3.6 $\pm$ 0.5  	&	2.88 		&	6 hours after MC \\
64 & 2002 Aug 24 	&  	010000  -  020000   	& 2.4 $\pm$ 0.5  	&	2.96 		&	Trail of ICME \\
65 & 2003 Oct  28 	&  	100000  -  110000   	& 4.3 $\pm$ 0.4  	&	2.39 		&	Trail of MC \\
66 & 2003 Oct  29 	&  	190000  -  200000   	& 5.7 $\pm$ 1.0  	&	4.15 		&	Trail of MC \\
67 & 2003 Nov 02 	&  	160000  -  170000   	& 3.3 $\pm$ 0.5	 	&	2.85		&	Trail of ICME \\
68 & 2005 Jan  17 	&  	090000  -  100000   	& NG  			&	2.72 		&	Trail of MC  \\
69 & 2005 Jan  20 	&  	053000  -  063000   	& 2.6 $\pm$ 0.3		&	2.31 		&	Trail of MC  \\
70 & 2006 Dec 13 	&  	010000  -  020000   	& 3.8 $\pm$ 0.6  	&	3.07 		&	Solar Wind \\
71 & 2012 May 17  	&  	013800  -  033000    	& 3.8 $\pm$ 0.6  	&	3.07 		& 	Inside MC 	 \\
\enddata

\tablecomments{Columns 1: The official GLE number, 2 and 3: the date, the start and end times of the GLE, 4: CME height (in solar radii, R$_\odot$) at the solar particle release time inferred derived by Reames (2009) using a velocity dispersion analyses based on near-Earth particle measurements, 5: CME height at GLE onset obtained by quadratic extrapolation (in solar radii, R$_\odot$) by Gopalswamy et al. (2012), 6: the state of the solar wind at GLE onset. Acronyms used: NG for not given, MC: Magnetic Cloud, ICME: Interplanetary Coronal Mass Ejection. \label{table1} }

\end{deluxetable*}

\indent Both the PFSS and MHD approach show that the shock progresses from a quasi-perpendicular shock to a quasi-parallel super-critical shock. The PFSS model suggests that a quasi-perpendicular shock forms around the SPR time derived by the velocity dispersion analysis (VD SPR). This result could support the idea that a quasi-perpendicular shock combined with a seed population of energetic particles may be more effective to accelerate particles to very high energies than a quasi-parallel shock (Tylka and Lee 2005). Sandroos and Vainio (2009) showed that the magnetic geometry of the ambient corona can have an effect of about one order of magnitude on the maximum energies reached by the process of diffusive-shock acceleration, and that for some field geometries 1 GeV energies are attainable, provided that seed particles with sufficiently high energies (100 keV) are available. The MHD model suggests that the quasi-perpendicular shock has already occured and changed to an oblique super-critical shock by the VD SPR time. It would be instructive to run particle acceleration models to see if the weak quasi-perpendicular shock pre-accelerated some particles that were eventually accelerated to very high energies by the super-critical quasi-parallel shock. 

\indent  The very significant rise in $M_{FM}$ at the tip of the streamer, where we infer that the HPS must generally form, occurs near the release time of very energetic particles inferred from Earth-based neutron monitors for both techniques. In-situ measurements of the HPS show that the ambient magnetic field drops by an order of magnitude and the density can increase by a factor of 4-5 (Winterhalter et al.1994), hence the Alfven speed decreases dramatically to favour the formation of super-critical shocks. Remote-sensing observations suggest this plasma sheet already exists in the corona forming above the tip of helmet streamers (Bavassano et al. 1997; Wang et al. 2009). GLEs are associated with CMEs that emerge within a few latitudinal degrees of the nominal footprint of the Parker spiral connecting the point of in-situ measurements (e.g. Gopalswamy et al. 2012), the present study would argue that a good connectivity is necessary to the shock regions crossing the vicinity of the tip of streamers and the associated neutral line. The curved field lines that form the streamer could also favor multiple field-line crossings of the shocks and efficient particle acceleration (Sandroos and Vainio 2006, 2009).\\

\indent  The HPS has a number of other interesting properties that make it a favorable location for strong particle energisation. Beside potentially boosting the $M_{FM}$, the high densities typically observed in white-light and measured in situ in the HPS would provide a localised increase in seed particles to be accelerated by the shock. The last decades of research has revealed how variable the HPS is, both spatially and temporally. The HPS contains signatures of small-scale transients that are released continually from the tip of streamers including small-scale magnetic flux ropes. The HPS is often entirely missed at 1AU when ICMEs, magnetic clouds and smaller transients 'replace' the standard HCS crossing, such as seen in Figure\ref{MAthetaBnPFSS}. These transients are formed inside or at the top of helmet streamers in regions where magnetic reconnection between oppositely directed field lines is very likely happening continually to form bundles of twisted magnetic fields or flux ropes. This continual transient activity leaves systematic signatures in the outflowing solar wind along the heliospheric neutral line (e.g. Rouillard et al. 2010, Plotnikov et al. 2016). The formation of these complex field topologies involves the closed magnetic field lines situated inside helmet streamers that are in more direct proximity to the flaring active region than open field lines from coronal holes.  We note that the tip of streamers may therefore provide an escape route for heavy ions and suprathermal particles that were previously confined to closed magnetic loops as either pre-energised particles by the quasi-perpendicular shock or by the concomittant flaring activity. Since the energetic particles move faster than the accelerator, they rapidly populate and scatter upstream of the forming shock in the open field region. These latter particles could also be an important population of seed particles for a prompt energisation by the mechanism of diffusive-shock acceleration (Tylka and Lee 2005). Even the MHD model used in this paper is not able to model such disruptions realistically, and therefore our derivation of the geometry of the shock, uncertain in these models, will be even more uncertain inside the HPS.

\indent Masson et al. (2012) analysed the near-Earth properties of the solar wind during 10 out of the 16 GLEs detected by neutron monitors since 1997. They showed that 7 of the GLE onsets occurred during disturbed solar wind conditions measured by ACE and Wind at 1AU including 2 very clear ICME passages. This frequent association is related to the finding made by Belov et al. (2009) that the accelerative and modulative efficiencies of solar storms are tightly correlated; CMEs followed by GLEs are associated with a high probability of a very large Forbusch decrease measured at Earth. We revisited the analysis by Masson et al. (2012) by (1) analysing the near-Earth solar wind conditions of the other official GLE events not listed in their paper, (2) considering in addition the suprathermal electrons measurements obtained by the ACE and Wind spacecraft, (3) the ICME list of Richardson and Cane (2009). The results are shown in column 6. As revealed by figure \ref{MAthetaBnPFSS}, the GLE event analysed in this paper occurred inside a clear magnetic cloud. Out of the 16 GLE events, we could confirm that the near-Earth environment was in the 'background' solar wind for two events only (GLEs 58 and 70).  \\

Using the full set of SECCHI observations, we demonstrated in Appendix  B that that the magnetic cloud measured in situ at the time of the GLE originated in a CME that erupted on 12 May 2012 from AR 11476. This agreement between the source longtiude of both the active region that produced the magnetic cloud measured in situ and the CME/GLE event of 17 May 2012 strengthens our argument that the near-Earth environment was magnetically connected, through a flux rope, to the coronal region that produced the shock, perhaps rooted in the direct vicinity of the active region. It remains to be demonstrated whether every ICME that occured during GLEs since 1997 erupted from the same region that produced the flare/CME responsible for the GLE event. For our event, the MHD simulation finds open field lines rooted near AR 11476, future simulations could investigate whether these open field lines were opened by the preceding CME activity. \\ 

\indent In light of the previous results, a limitation of our study resides in the rather static treatment of the background coronal magnetic field. The eruption of CMEs from the same active region days prior to the event of insterest in this paper may have induced time-dependent effects that are missed out by the PFSS and MHD model used in present study. Previous observational studies combined with numerical models of the coronal field have investigated the topological changes induced by CME eruption. They demonstrate that CMEs (1) open closed field lines that previously formed the streamer's helmet base (e.g. Fainshtein et al. 1998), (2) generate additional white-light rays in the trailing part of CMEs that appear for several hours (Kahler and Hundhausen 1992; Webb et al. 2003), (3) produce transient coronal holes in less than 1 hour that disappear in 1–2 days (e.g. de Toma et al. 2005). The event of 2012 May 12 occurred five days before the event studied here and these transient structures had faded by the 2012 May 17. The white-light rays in particular were clearly visible in the plane of the sky from STA and had largely faded away from the camera that same day. The MHD model suggested that some of the field lines connected to the plasma sheet were rooted in the AR11476 and could be remnants of this preceding CME activity.  \\

\begin{figure}
\begin{center}
\includegraphics[angle=0,scale=.37]{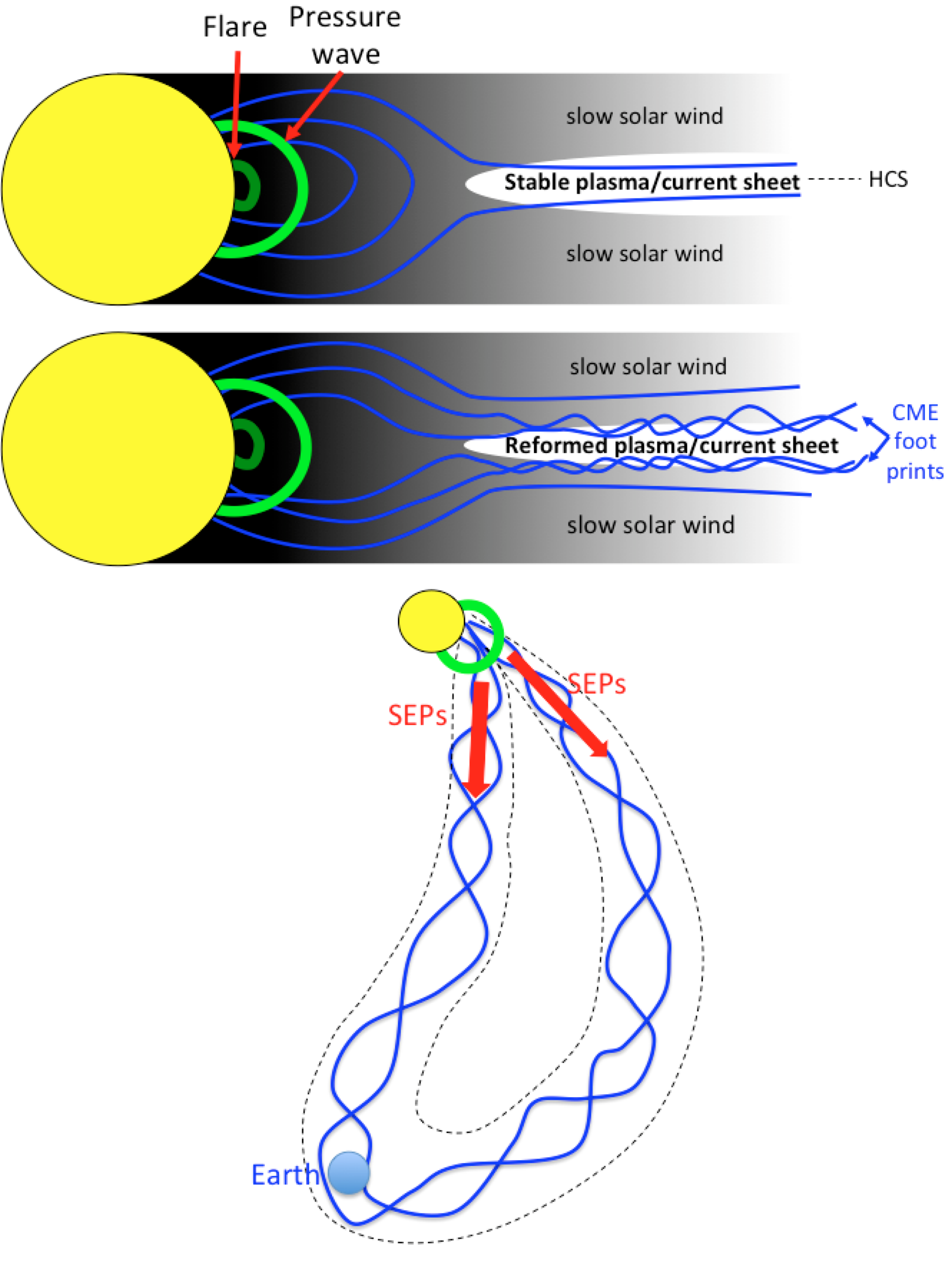}
\caption{The top two schematics are views onto helmet streamers with magnetic field lines shown in blue. The location where oppositely directed field lines meet is the heliospheric plasma and current sheet. The fall off of the coronal fast-mode speed with distance is shown as a fading black color illustrating the abrupt drop usually seen inside the plasma sheet. In green, we show the relative locations of the flare and the pressure wave that forms around the CME acting as a piston. Two scenarii are sketched during 'quiet' (top image) and more 'disrupted' (middle image) coronal conditions during the reformation of the plasma sheet over the several days that follow the eruption of a CME. Bottom panel: a schematic of the interplanetary conditions during the GLE event. A magnetic cloud with closed field topology is connecting the coronal shock with the near-Earth environment, channeling particles to 1AU. }
\label{HPSFR}
\end{center}
\end{figure}

\indent The presence of a magnetic cloud linking the shock to the near-Earth environment could change the magnetic connectivity of the near-Earth environment with the coronal plasma sheet. As stated above the helmet streamers retrieve an equilibrium configuration at most a few tens of hours after the eruption of a CME, during this process it is unclear where the magnetic footpoints of CMEs end up connected to in the low corona but presumably they will be part of the new equilibrium configuration found by the streamer and its reformed plasma sheet. The top two schematics of Figure \ref{HPSFR} present an illustration of the relation between the emerging shock and the plasma sheet with no CME activity prior to the event and for a disturbed plasma sheet reaching a new equilibrium but threaded by magnetic field remnants of prior CME activity rooted near the active region.   \\

\indent The nearly systematic association between the occurrence of GLEs and the passage near Earth of ICMEs will also change the likelihood of being connected with the accelerator near the Sun. We note that contrary to a single Parker spiral, a magnetic flux rope occupies a very large volume of the interplanetary medium; this will increase the probability that the near-Earth environment (or any point inside the flux rope) becomes magnetically connected with the coronal region producing very high-energy particles. If the neutral line is indeed a favorable but spatially limited region of particle acceleration, the presence of a large-scale magnetic flux rope (or another complex magnetic field structure) will increase the chances of being magnetically connected with that narrow region. This is illustrated in the bottom schematic of Figure \ref{HPSFR}. The magnetic cloud passing over the Earth on 2012 May 17 transports counter-streaming suprathermal electrons just before the onset of the GLE event. In addition, the magnetic connectivity to the solar corona was therefore  occurring at both ends of the flux rope and SEPs. Strong beams of counter-streaming electrons are often measured after the onset of GLE events but these are likely associated with back-propagating electrons due to the ICME acting as a magnetic mirror and an associated higher flux of electron due to the GLE and a higher level of scattering due to the ICME magnetic fields.  \\

\section{Conclusion:}

Assuming that the particle accelerator is situated near the shock-sheath system, the delay typically seen between the flare and solar particle release times should depend initially on the time necessary:

\begin{itemize}

\item   for the pressure wave to steepen into a shock: the formation processes of the shock will depend on the 3-D expansion speed of the driver gas and the spatial variations of the characteristic speed of the ambient medium in which it is propagating, 

\item  for the shock to propagate longitudinally: this is particularly true during the progression through predominantly radial magnetic fields since cross-field diffusion is much weaker than field-aligned diffusion.

\end{itemize}

The use of different magnetic models points to the considerable uncertainties that are faced when attempting to derive the topology of the background magnetic field through which coronal shocks propagate. However our approach has revealed that, regardless of the model used, a shock has formed at the time of the onset of the type II burst and a super-critical shock has formed at the release time of high-energy particles.

An alternative hypothesis not investigated here is of course that particles are accelerated in the solar flare. The delayed GLE onset would then be interpreted as the time required for the closed loops that drive the expansion of the piston and also channel the flare particles, to reconnect with the open magnetic field lines that are connected with the spacecraft measuring the GLE. This mechanism was investigated numerically by Masson et al. (2013). For the event analysed here, the reconnection process would occur between the erupting piston and the magnetic field lines of the magnetic cloud measured in situ. A delayed onset could only occur here if the magnetic field lines of the magnetic cloud are initially topologically distinct to the flaring loops or the erupting piston.\\

\indent We are currently repeating the analysis presented in the present paper on other events measured by near 1AU orbiting spacecraft with the hope to decipher the nature of the particle accelerator. Clearly the presence of additional spacecraft situated closer to the Sun (Solar Probe+) and outside of the ecliptic plane (Solar Orbiter) should provide (1) radically better timing of particle onsets than inferred by in-situ measurements made near 1AU and (2) unprecedented views from outside the ecliptic plane to disentangle more easily the different delays in particle onsets.\\

\acknowledgments
We acknowledge usage of the tools made available by the plasma physics data center (Centre de Données de la Physique des Plasmas; CDPP; http://cdpp.eu/), the Virtual Solar Observatory (VSO; $http://sdac.virtualsolar.org$), the Multi Experiment Data $\&$ Operation Center (MEDOC; $https://idoc.ias.u-psud.fr/MEDOC$), the French space agency (Centre National des Etudes Spatiales; CNES; https://cnes.fr/fr) and the space weather team in Toulouse (Solar-Terrestrial Observations and Modelling Service; STORMS; https://stormsweb.irap.omp.eu/). This includes the data mining tools AMDA (http://amda.cdpp.eu/) and CLWEB (clweb.cesr.fr/) and the propagation tool (http://propagationtool.cdpp.eu). R.F.P. and I.P. acknowledge financial support from the HELCATS project under the FP7 EU contract number 606692. R.V. acknowledges the financial support from the HESPERIA project under the EU/H2020 contract number 637324. A. W. acknowledges  the support by DLR under grant No. 50 QL 0001. A.P.R. acknowledges funding from CNES and the Leibniz Institute f\"{u}r Astrophysik Potsdam (AIP) to visit AIP and to collaborate with A.W. and G.M. on the present project. The \emph{STEREO} \emph{SECCHI} data are produced by a consortium of \emph{RAL} (UK), \emph{NRL} (USA), \emph{LMSAL} (USA), \emph{GSFC} (USA), \emph{MPS} (Germany), \emph{CSL} (Belgium), \emph{IOTA} (France) and \emph{IAS} (France). The \emph{ACE} data were obtained from the \emph{ACE} science center. The WIND data were obtained from the \emph{Space Physics Data Facility}. 





\appendix
\section{Deriving the solar particle release time:}
\begin{figure}
\begin{center}
\includegraphics[angle=0,scale=.3]{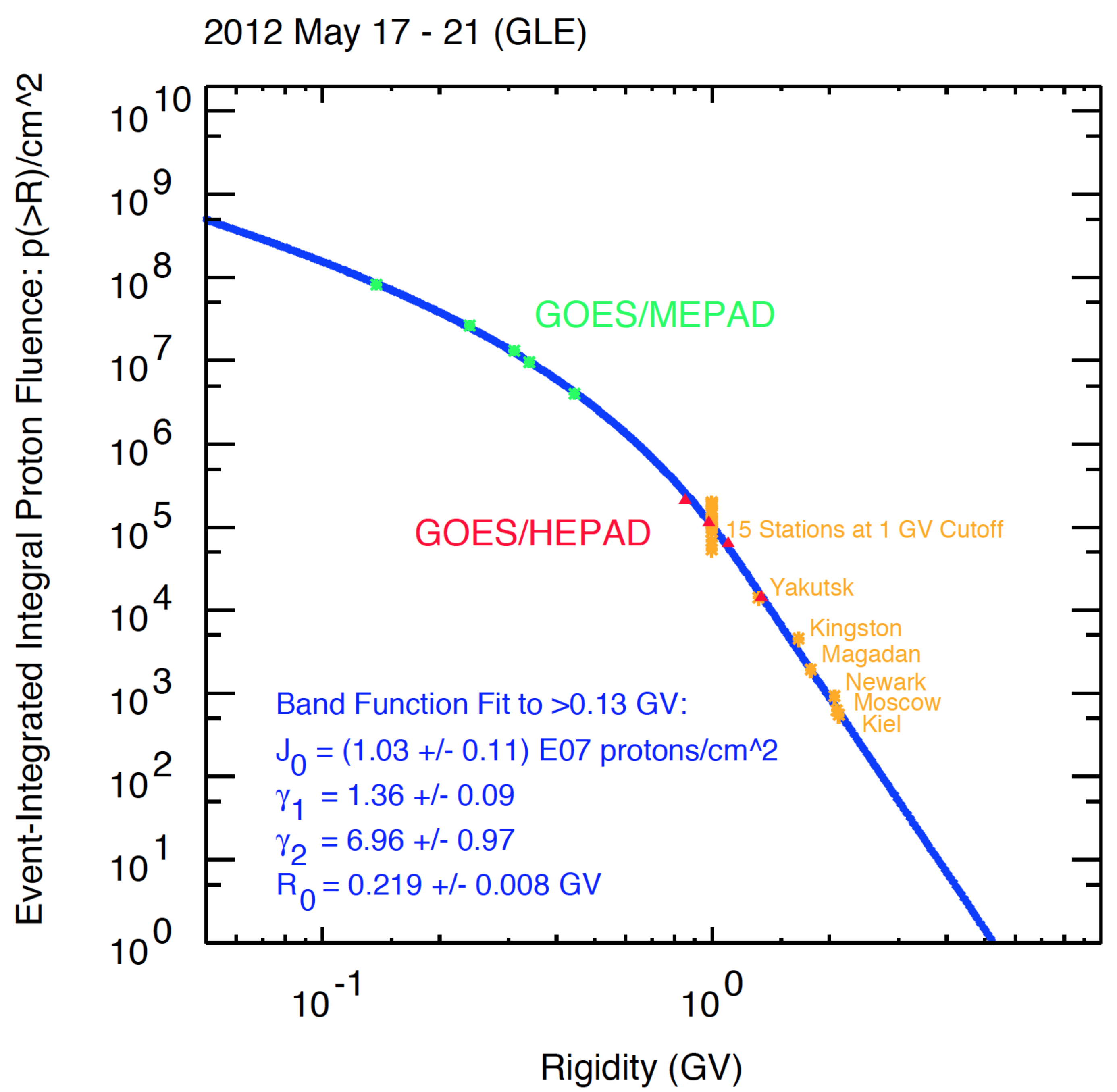}
\caption{A event-integrated spectrum derived from the GOES/MEPAD, GOES/HEPAD proton data combined with an analysis of the neutron monitor data obtained by 15 stations at 1GV cutoff as well as six other neutron monitors with higher rigidity cutoffs. }
\label{SPEC}
\end{center}
\end{figure}

We here derive a velocity dispersion analysis based on the arrival times of different particles with different energies measured near Earth. A velocity dispersion analysis is obtained by plotting the onset times of particle flux increases versus the reciprocal of the relativistic beta value ($\beta^{-1} = (v/c)^{-1}$) for different particle energies (Krucker et al. 1999; Tylka et al. 2003). A linear fit on such a scatter plot determines the initial solar particle release (SPR) time, as the intercept, and the path length followed by the particles between the Sun and L1, as the slope. The available proton data used for this analysis was obtained by the ERNE (101-131MeV) and COSTEP instruments onboard SOHO (25-60MeV), the helium data (1.65$-$9.64 MeV/nucleon) was obtained by the LEMT instrument onboard Wind and the neutron data used is from the Oulu neutron monitor. The geomagnetic cutoff at Oulu is 0.9GV (360MeV) but the Oulu response is governed by the atmospheric cutoff, which is about 1GV (435 MeV). Moreover the first arriving protons are likely to be the highest energy ones. The event-integrated proton spectrum shown in Figure \ref{SPEC} shows that protons were detected up to at least 2.1GV (1385MeV). Thus, when calculating the 1/beta values for the onset analysis, the energies that should be considered are considerably higher than used in Gopalswamy et al. (2013).\\

\begin{figure}
\begin{center}
\includegraphics[angle=0,scale=.27]{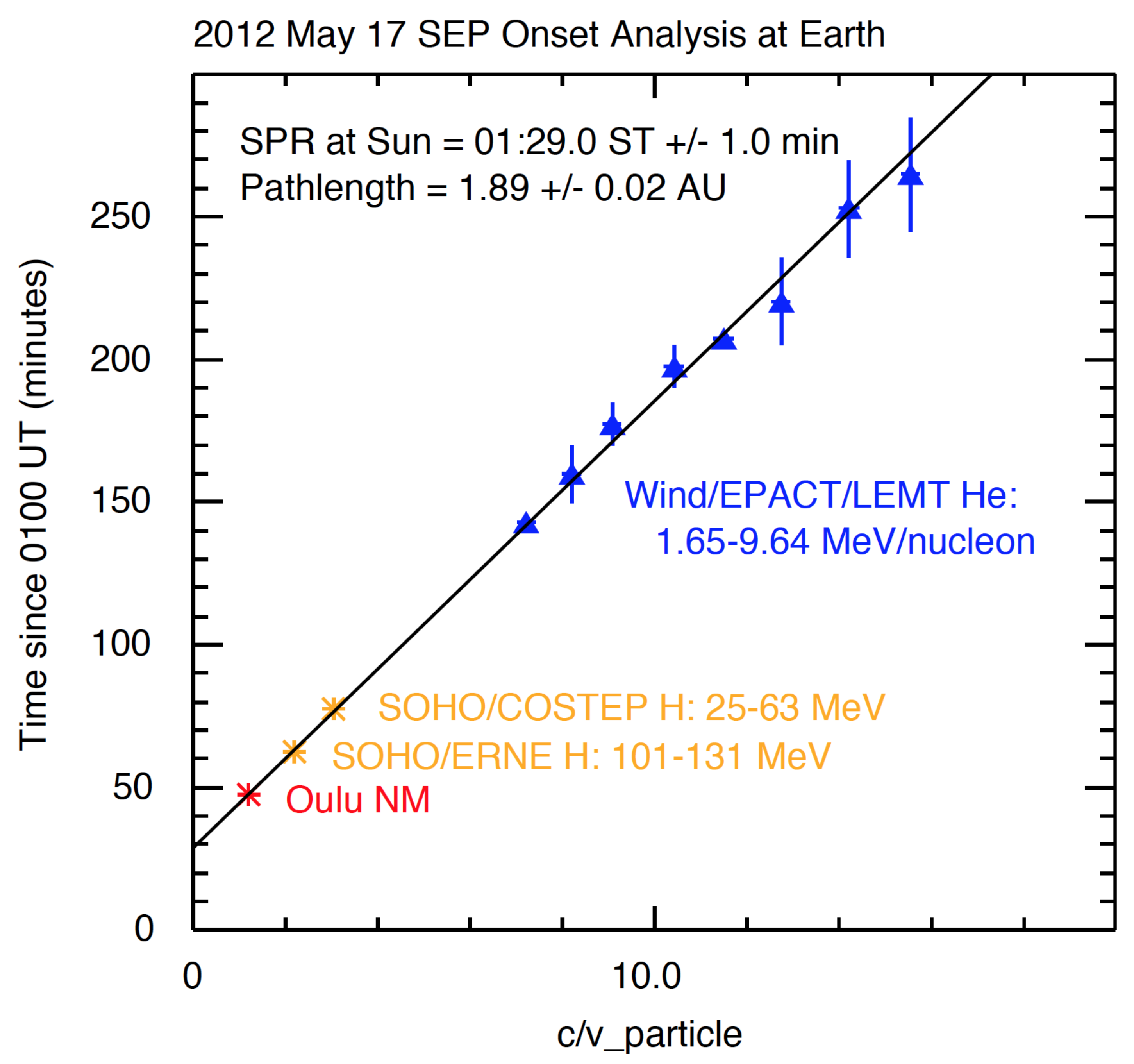}
\caption{Top panel: a velocity dispersion analysis based on the measurements
of the onset of light (protons), and heavy ions (oxygen, 4helium,
and iron). The proton data was recorded by the SOHO ERNE and COSTEP instruments. The Helium data was recorded by the LEMT instrument onboard Wind, the onset of the Ground Level Enhancement was obtained by the Oulu neutron monitor (rigidity cutoff: 0.9GV). }
\label{SPR}
\end{center}
\end{figure}

Figure \ref{SPR} presents this velocity dispersion analysis. The estimated release time of the particles is 01:29$\pm$1minUT and the pathlength is 1.89 $\pm$0.02 AU.  To compare with electromagnetic radiation we must add 8.41 minutes corresponding to the time for light to travel to the Earth so 01:37:20UT. The release time derived by this analysis is just under 4 minutes earlier than the release time derived by Gopalswamy et al. (2013) using a simpler approach. This earlier time shift is in part related with the rather long pathlength of 1.89 AU derived using the velocity dispersion. Numerical simulations of particle transport in the interplanetary medium combined with the effect of self-generated waves upstream of the shock show that pathlengths derived from velocity dispersion analyses can be slightly longer than the simple Parker spiral because particles scatter off irregularities (see Appendix A of Rouillard et al. 2012). However Litunen and Vanio (2004), who extend the dispersion analysis to much lower energies and considered a simpler scattering model, found that larger distortions in the pathlength are possible. \\

\indent Another effect that should be considered for the present event, is that the SEP was detected during the passage of a magnetic cloud at Earth. Hence particle were streaming not along simple Parker spirals but rather along helical magnetic field lines that form the magnetic cloud. This could increase the pathlength travelled by energetic particles during their transport to 1AU. The standard picture of a flux rope is a helical (poloidal) magnetic field line wound around a straighter toroidal magnetic field suggests that magnetic field lines nested near the boundary of magnetic flux ropes should be several times longer than near the center. A recent study by Kahler et al. (2011) investigated whether the pathlength of energetic electrons detected on the boundary of magnetic clouds differed from the pathlengths measured near the center of magnetic cloud and in the quiet solar wind. They considered different magnetic models for the magnetic cloud and found generally poor correlations between the computed electron path lengths and the model field line lengths.\\

\indent The onset of the May 17 SEP event occurred right at the center of the magnetic cloud where the toroidal component of the magnetic field is dominant. There is currently no accepted magnetic field topology for magnetic flux ropes particularly of the legs magnetically connected to the Sun. The simplest, but likely inaccurate, picture of a CME flux rope is that of a straight toroidal flux rope with, at the time of impact of the structure at 1AU, half the Sun-Earth distance as its major radius. Such a structure would enclose quasi-circular field lines passing along the center of the toroid with lengths of about $\pi/2=1.57$AU. This distance is slightly less than the pathlength derived by the velocity dispersion analysis (Figures \ref{SPR}). Nevertheless if one considers the fact that particle diffusion in the corona and in the interplanetary medium may add a factor 0.1 to the travel pathlength (Rouillard et al. 2012), we get a pathlength of $\sim$1.7AU which is close to the pathlength derived by the velocity dispersion analysis.  For a nearly horizontal flux rope, such as the one measured in situ during the May 17 event, such an idealised set of toroidal field lines would connect the Earth to the eastern and western limb of the Sun and therefore to the vicinity of AR 11476 that produced the shock analysed in this paper. 

\section{Deriving the state of the IP medium:}

In this section we present an analysis of solar wind conditions over the days that preceded the GLE event in order to determine how the near-Earth environment is connected with the low corona at GLE onset. Since we know that a clear magnetic cloud passed over the Wind and ACE spacecraft at the time of the GLE, we can make the reasonable hypotheses that (1) a CME that erupted several days early propagated along a trajectory close to the Sun-Earth line and that this CME is magnetically connected with the particle accelerator that produced the GLE. This appendix seeks to test those assumptions by using heliospheric imagery in order to gain some insight on the longitudinal variability of solar wind conditions right before and during the GLE . The orbital configuration of the STEREO spacecraft was such that the heliospheric imagers onboard STA and STB were continuously monitoring plasma outflows along the Sun-Earth line at the time and as demonstrated in previous case studies were ideally suited to study transient activity continually driven along specific longitudes in the corona.  \\

\begin{figure}
\begin{center}
\includegraphics[angle=0,scale=.33]{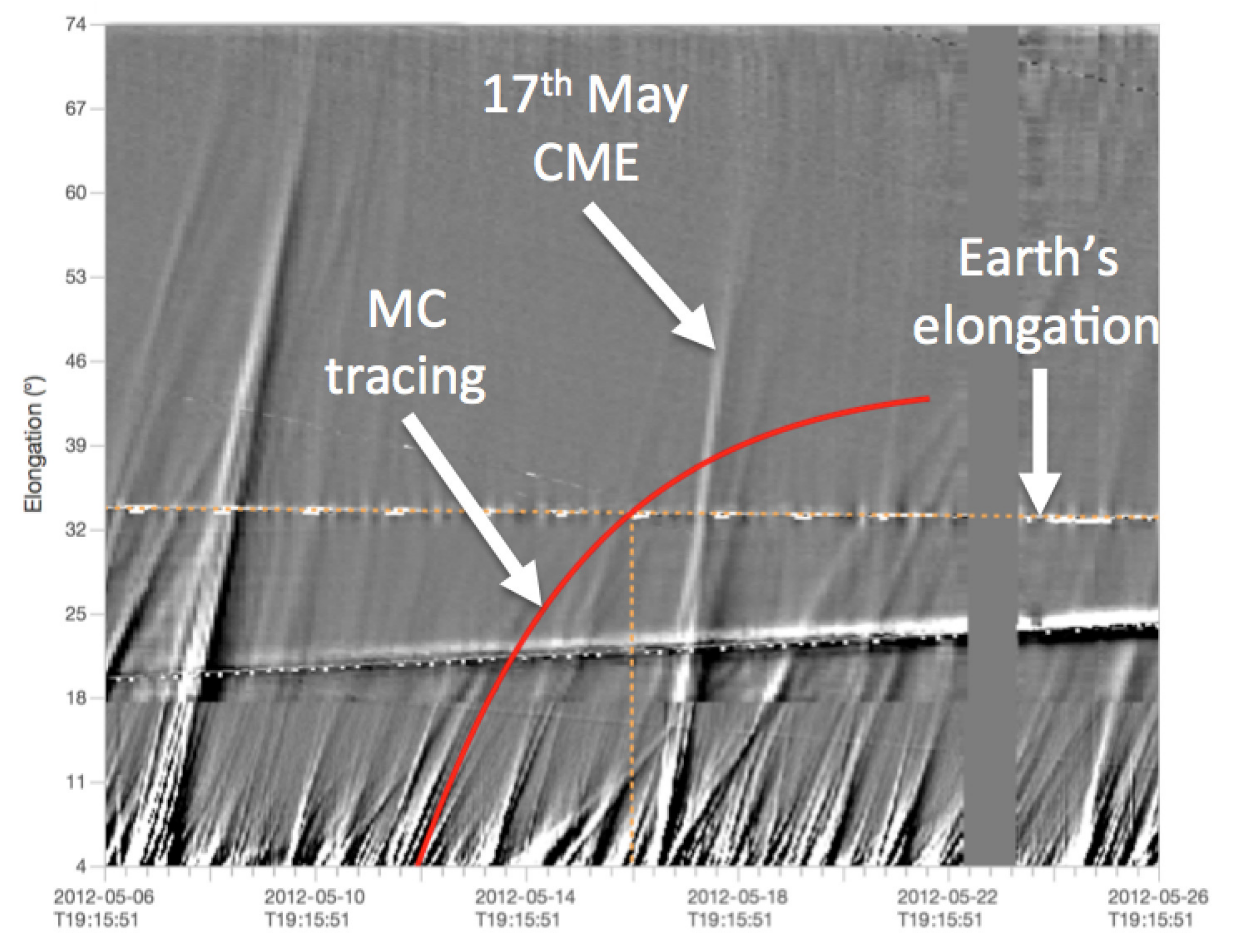}
\caption{Panel a: A J-map derived from heliospheric imaging made by STA showing the state of the IP medium between the 7$^{th}$ and 27$^{th}$ of May 2012. Each track on these J-maps corresponds to a density structure moving radially outward from the Sun and leaving a strong signature in the WL imagers. The Earth's elongation is shown as a horizontal yellow dotted line. The time of passage of the MC detected at Earth at the start of the 17$^{th}$ of May 2012 (c.f. Figure \ref{INSITUSEP})  is shown as a vertical dotted line intersecting the elongation of the Earth, since the GLE occurred at exactly that time, this vertical line also marks the onset of the GLE event. The apparent elongation variation of the MC is shown as the red track on this map superposed on a clear track seen in the J-map. This figure was produced using the IRAP propagation tool (propagationtool.cdpp.eu) configured in the 'radial/Carrington/In situ' mode.}
\label{JMAP_12MayCME} 
\end{center}
\end{figure}

To track individual features precisely in the field of view of the heliospheric imagers, maps of brightness variations are usually created by extracting a band of pixels situated along a solar radial corresponding to a constant position angle (PA) and displaying this band as a function of elongation (Y-axis) and time (X-axis; (Sheeley et al., 1999, 2008b,a; Davies et al., 2009)). To track plasma propagating towards a spacecraft situated in the ecliptic plane, this PA is left to vary slowly with time with the orbital motion of the imager so that the band of pixels extracted to form the J-map tracks systematically brightness variations along the ecliptic plane. Such a J-map is shown in Figure \ref{JMAP_12MayCME}. The angular range (vertical axis) of the J-map goes from 4 to 74$^{\circ}$. This range includes the elongation of Earth, hence STA was at the time imaging plasma flowing between the Sun and Earth.\\ 

\indent Using the in-situ speed of the cloud, we follow the technique presented in Rouillard et al. (2011b) to derive the apparent trajectory that a CME impacting a specific probe. We assume that the CME is a point-like structure (so-called 'fixed-phi approach' e.g. Rouillard et al. 2008)  and that the CME has a speed of 350 km/s measured in situ inside the MC (c.f. Figure \ref{INSITUSEP}). The inferred CME trajectory leaves the red track in the J-map shown in Figure \ref{JMAP_12MayCME}, it matches very closely the track of a CME that erupted at around 00UT on 12 May 2012. The procedure to superpose apparent trajectories on J-maps is integrated in a web-based interface developed by the Research Institute for Astropysics and Planetology (IRAP) in Toulouse and named the 'propagation tool' (propagationtool.cdpp.eu). Assuming that no CME deflection occurred in the low corona, the tool puts the source location of the CME at the same Carrington longitude as the active region that produced the CME of the GLE event (AR11476).\\

\begin{figure}
\begin{center}
\includegraphics[angle=0,scale=.75]{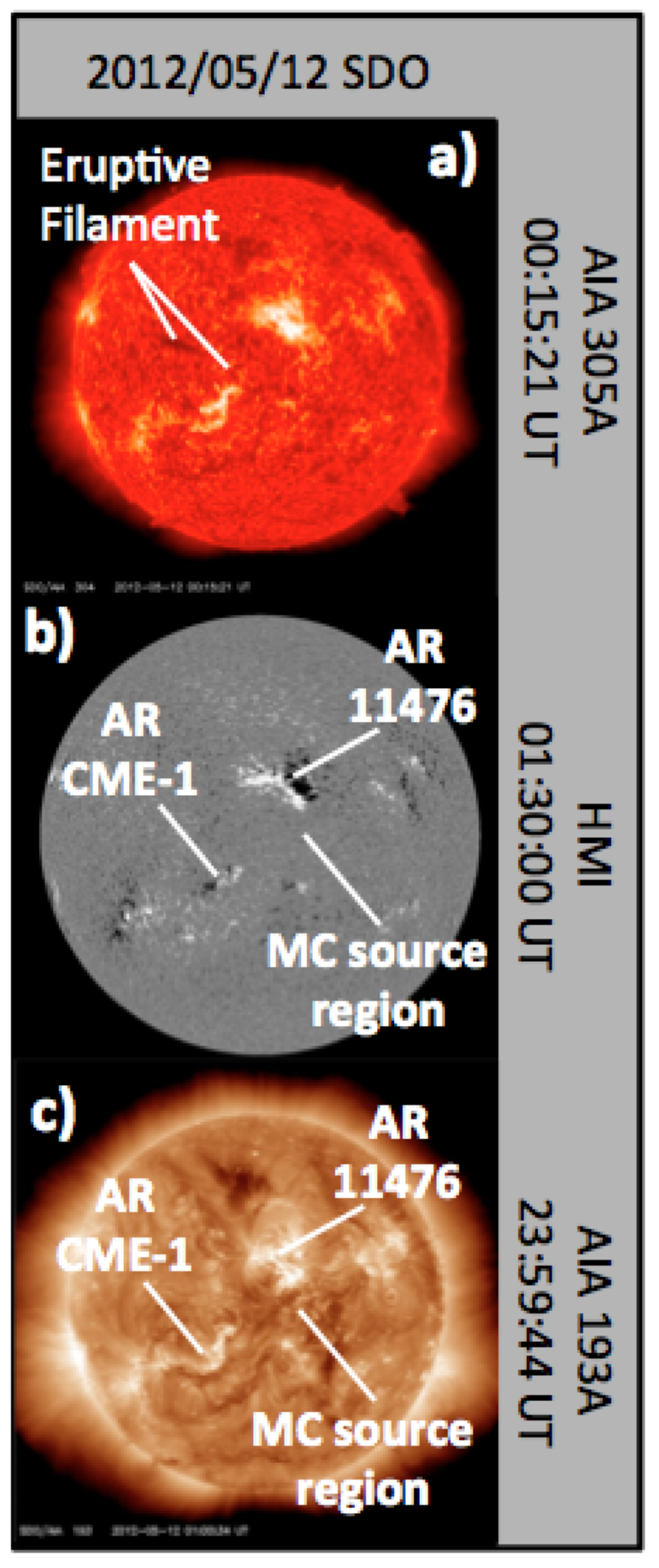}
\caption{Observations of the photosphere and corona made by the SDO on 12$^{th}$ May 2012. Panel a: AIA observations of the corona in 305 \AA at the time of the eruptive filament and estimated launch time of the magnetic cloud measured in situ on 2012 May 17. Panel b: a magnetogram obtained by the HMI instrument showing the active regions (AR 11476) that produced the 17 May 2012 CME/GLE events and the active region (AR MC) that produced the magnetic cloud measured in situ on 17 May 2012. Panel c: observations of the corona made in 193\AA several hours after the 12 May 2012 CME event showing the coronal holes that gradually formed following that eruption.}
\label{SDO_12MayCME}
\end{center}
\end{figure}
\indent Inspection of the EUV imager reveals that at the same time as the release of the small CME from AR11476, an off-limb prominence eruption was seen by STA and STB EUVI in 304 \AA with a coincident on-disk filament disappearance by SDO AIA in 305\AA. Figure \ref{SDO_12MayCME} shows the filament at 00:15UT just before it disappeared from SDO AIA images. The source location of the combined large prominence/CME eruption occurs 20 degrees eastward of disk center (Figure \ref{SDO_12MayCME}a) in contrast to the CME from AR11476 that erupted 5 degrees westward of disk center. The location of the different source locations are indicated on the HMI magnetogram shown in Figure \ref{SDO_12MayCME}b and on the EUV image taken at 193\AA in Figure \ref{SDO_12MayCME}c. \\

\begin{figure*}
\begin{center}
\includegraphics[angle=0,scale=.52]{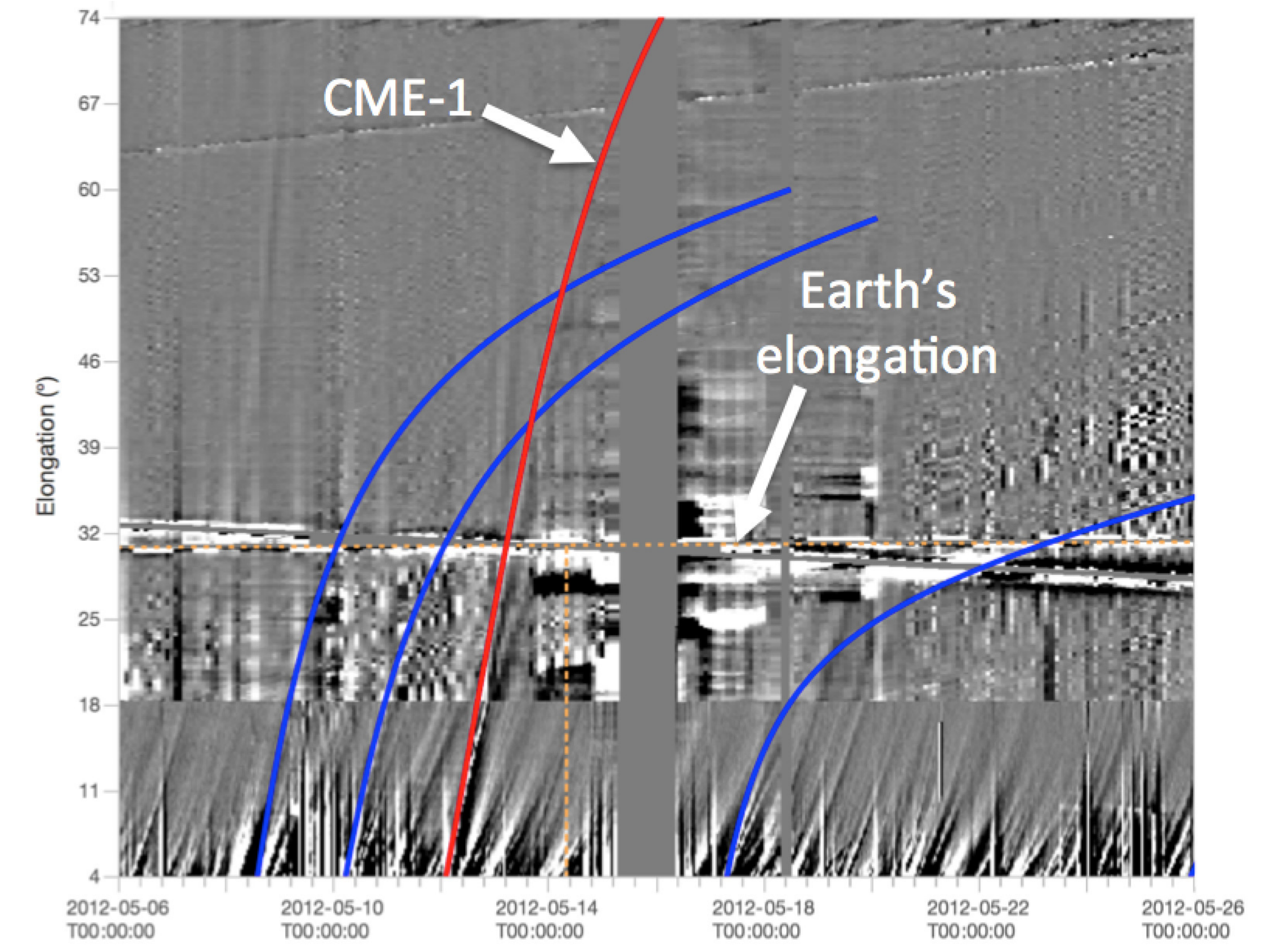}
\caption{A J-map in the same format as Figure \ref{JMAP_12MayCME} but derived with STB observations. The blue and red tracks show the apparent trajectory of different CMEs provided by the catalogues produced by the HELCATS FP7 project. The red track marks the trajectory of the CME-1 that erupted on 12 May 2012 to the East of the active region that produced the SEP event analysed in the paper. This analysis shows that only the flank of CME-1 would have impacted the Earth before the SEP event on 14 May 2012.}
\label{JMAP_ARCME1}
\end{center}
\end{figure*}
\indent The HELCATS project has recently published various catalogs of CMEs observed in the SECCHI cameras onboard STEREO and has made them available on the project website  (http://www.helcats-fp7.eu/products.html). The catalog includes for each CME, various estimates of their trajectories and speeds using the fixed-phi (Rouillard et al. 2008), harmonic mean technique (Lugaz et al. 2009) and self-similar expanding techniques (Davies et al. 2009). The combined prominence/CME eruption left a strong signature in the heliospheric images taken by the STB spacecraft shown in Figure \ref{JMAP_ARCME1}. The various estimates of this CME's trajectory listed in the HELCATS CME catalogue point to a direction of propagation along an averaged Carrington longitude of $150\pm25^\circ$. While the fixed-point technique predicts no impact of CME-1 at Earth (unless it extended over a broad 90$^\circ$ in longitude), the self-similar technique predicts an impact of the far Western flank of the CME at 19UT on 14 May 2012 (for a more reasonable spread in longitude of 45$^\circ$). This corresponds to the arrival of the ICME detected by ACE and Wind before the magnetic cloud (Figure \ref{INSITUSEP}). Hence although two CMEs erupted at the same time from the visible disk, their arrival times at 1AU were very different due to the their different speeds and directions of propagation. Both the ICME and the magnetic cloud were roughly connected with the region of the streamer  and channeled particles accelerated by the shock on 17 May 2012.  \\

\indent We also checked whether a Corotating Interaction Region (CIR) was passing in the field of view of the imagers by analysing the CIR catalogue developed by the  Heliospheric Cataloguing, Analysis and Techniques Service (HECALTS). Consideration of the CIR catalogue derived from STA images (Plotnikov et al. 2016)  reveals that no CIR was passing near Earth during the GLE event. \\



\section*{References}
\noindent
Adriani, O., et al. 2015a, \apj, 801 L3, doi:$10.1088/2041-8205/801/1/L3$\\
Arden, W. M., Norton, A. A.; Sun, X. 2014, \jgr, 119, 3, 1476\\	
Aschwanden, M.J., Schrijver, C.J., Alexander, D. 2001,  \apj, 550, 2, 1036-1050\\
Bavassano, B., Woo, R.; Bruno, R. 1997, \grl, 24, 13, 1655-1658\\
Belov, A. 2009, Uni. Helio.l Proc., Proceed. of the Int. Astron. Union, IAU Symposium, 257, 439-450\\
Bemporad, A., Susino, R., Lapenta, G. 2014, Astrophys. J., 784, 2, 102\\
Cane, H.V., von Rosenvinge, T.T., Cohen, C.M.S., Mewaldt, R.A. 2003, Geophys. Res. Lett., 30, 12, CiteID 8017\\
Billings, E. 1996, A guide to the Solar Corona, New York: Academic Press, |c, 1966\\
Cane, H.V. \& Richardson, I.G. 2003, \jgr, 108, A4, SSH 6-1, CiteID 1156, $DOI 10.1029/2002JA009817$\\	
Cerutti, B. et al. 2015, \mnras, 448, 606\\
Cohen, C.M.S et al. 2014, \apj., 793, 1, 35, 10\\
Crooker, N. U. et al. 2004, \jgr, 109, A3, $CiteID A03107$\\
Crooker, N. U. et al. 1996, \jgr, 101, Issue A2, p. 2467-2474\\
De Toma, G.  2005, \apj 621:1109–1120\\
Edmiston, J. P., \& Kennel, C.F. 1984, J. Plasma Phys., 32, 429\\
Fainshtein, V.G.; Rudenko, G.V.; Grechnev, V.V. 1998, Sol. Phys., 181, 1, 133-158\\
Gopalswamy, N., et al. 2012, \ssr, 171, 1-4, 23-60\\	
Gopalswamy, N.,  et al. 2013, \apj, 765, 2, article id. L30, 5\\
Harrison et al. 2016, in preparation\\
Howard, R.A. et al. 2008, \ssr, 136, 67\\
Hundhausen, A.J. 1972, Coronal Expansion and Solar Wind, Springer, Berlin Heidelberg New York\\
Kahler, S. W.; Hundhausen, A. J. 1992, \jgr, 97, 1619-1631\\
Klassen, A., Aurass, H., Mann, G., Thompson, B.J. 2000, A \& A Suppl., 141, 357\\
Krucker, S., Larson, D. E., Lin, R. P., \& Thompson, B. J. 1999, ApJ, 519, 864\\
Kozarev, K. et al. 2011, \apj, 733, 2, article id. L25\\
Kozarev, K. et al. 2013, \apj, 778, 1, 43, 13\\
Kozarev, K. et al. 2015, \apj, 799, 2, article id. 167, 10\\
Kwon, R.Y., et al. 2015, \apj, 799, Issue 2, article id. 29, 5\\
Hundhausen, A.J. 1972, Coronal Expansion and Solar Wind, Springer, Berlin Heidelberg New York\\
Lario, D., et al. 2014, \apj, 797, 16\\
Leblanc, Y., Dulk, G. A., Bougeret, J.-L. 1998, Sol.Phys., 183, 1, 165-180\\
Lemen, J., et al. 2012, Sol. Phys., 275, 1-2, 17-40\\
Linker, J. A. et al. 1999, \jgr, 104, A5, 9809-9830\\
Lionello, R. et al. 2009, ApJ, 690, 902\\
Luhmann, J.G., Curtis, D.W., Schroeder, P., McCauley, J., Lin, R.P., Larson, D.E., et al. 2008, \ssr, 136, 117\\
Manchester, W., et al. 2008, ApJ, 684, 2, 1448, 1460\\
Mann, G. 1995, in Coronal Magnetic Energy Release, ed. A. O. Benz, \& A. Kruger, Lecture Notes in Physics, 183\\
Mann, G. et al. 1999, Astron. \& Astrophys., 348, 614-620\\
Mann, G. et al.  2003,  A\&A 400, 329\\
Marcowith, A., et al. 2016, Rep. on Prog. in Phys., 79, 4, 046901 2016\\
Mason, G.M., Gold, R.E., Krimigis, S.M., Mazur, J.E., Andrews, G.B., Daley, K.A., et al. 1998, \ssr, 86, 409\\	
Mason, G.M., Mazur, J.E., Dwyer, J.R. 1999, ApJ, 525, L133\\	
Masson, S., Antiochos, S. K., DeVore, C. R., \apj, 771,  2, 82, 15, 2013\\
Masson, S.,  et al., \aa, 538, A32, 14, 2012\\
Mewaldt, R.A. et al. 2008, \ssr, 136, 1-4, 285-362\\
Moestl, C. et al. 2009, \apj, 705, 2, L180-L185\\
Muller-Mellin, R. et al. 1995, Sol. Phys., 162, 483\\
Ontiveros, V. \& Vourlidas, A. 2009, \apj, 693, 267\\
Tylka, A.J., Lee, M.A. 2006, ApJ, 646, 1319\\
Patsourakos, S., Vourlidas, A. 2009, ApJ Lett., 700, L182\\
Picozza, P. et al. 2007, Astroparticle Physics, 27, 4, 296-315\\
Plotnikov, I., Rouillard, A.~P., Davies, J.~A., et al.\ 2016, \solphys, 291, 1853 \\
Poomvises, W. et al., O. 2012, \apj, 758, 2, 118, 6\\
Reames, D.V. 2009, \apj, 706, Issue 1, 844-850\\
Richardson, I.G. and Cane, H.V. 2010, Sol. Phys., 264, 1, 189-237\\
Rouillard, A.P. et al.  2007, \jgr, Sol. Phys., 112, A5, CiteID A05103\\
Rouillard, A. P. et al. 2008, \grl, 35, 10, CiteID L10110\\
Rouillard, A. P. et al. 2010, \jgr, 115, A4, CiteID A04103\\	
Rouillard, A.P., et al. 2011a, \apj, 735, 1, article id. 7 \\
Rouillard, A.P.  2011b, J. Atmos. Solar Terres.,Phys., 73, 10, 1201-1213\\
Rouillard, A. P., et al. 2011c, \apj, 734, 1, 7, 10\\ 
Sandroos, A., \& Vainio, R. 2006, A\&A,455, 685–695\\
Sandroos, A., \& Vainio, R. 2009, A\&A, 507, L21-L24\\
Salas-Matamoros, C., Klein, K.-L., Rouillard, A.P. 2016, Astron. \& Astrophys.., 590, A135, 15\\
Schwartz, S., 1998,  ESA/ISSI, Vol. 1. ISBN 1608-280X, 249-270\\
Scherrer, P. H. et al. 2012, Sol. Phys., 275, 1-2, 207-227\\
Smith, E.J.; Balogh, A. 1995, \grl, 22, 23, 3317-3320\\
Thompson, B. et al. 1998, GRL, 25, 14, 2465\\
Tylka, A.J. et al. 2003, Proc. 28th ICRC. (Trukuba), 3305\\
Vourlidas, A. \& Ontiveros, V. 2009, AIP Conf. Proceed., 1183, 139\\
Vourlidas, A. et al. 2013, Sol. Phys., 284, 1, 179\\
Wang, Y.-M.; Sheeley, N. R., Jr. 1992, \apj, Part 1, 392, 1, 310-319\\
Wang, Y.-M. et al. 2000, \jgr, 105, A11, 25133-25142\\
Wang, Y.-M. 2009, \ssr, 144, 1-4, 383-399\\
Wang, Y.-M. 2010,  \apj, 715, 1, article id. 39-50\\
Warmuth, A., Mann, G. 2005, Astronomy and Astrophysics, 435, 1123\\
Warmuth, A. 2015, Liv. Rev. Sol. Phys., 12, 3, $doi:10.1007/lrsp-2015-3$\\ 
Webb, D.F., Burkepile, J., Forbes, T.G., Riley, P. 2003, \jgr, 108, A12, SSH 6-1\\
Winterhalter, D.; Smith, E. J.; Burton, M. E.; Murphy, N.; McComas, D. J. 1994, \jgr , 99, A4,  6667-6680\\
Zucca, P., Carley, E.P., Bloomfield, D.S., Gallagher, P.T. 2014, A \& A, 564, A47, 9\\


\end{document}